%% file: main.tex
\documentclass[sigconf]{acmart}

\AtBeginDocument{%
  \providecommand\BibTeX{{%
    \normalfont B\kern-0.5em{\scshape i\kern-0.25em b}\kern-0.8em\TeX}}}

\usepackage{grffile}

\usepackage[normalem]{ulem}
\usepackage{amsmath}
\usepackage{hyperref}
\usepackage{tikz}
\usepackage{comment}
\usetikzlibrary{shapes,decorations,shapes.geometric,calc,arrows,matrix,shapes.arrows}
\usepackage{siunitx}
\usepackage{booktabs}
\usepackage{threeparttable}
\usepackage{subcaption}
\usepackage{array}
\usepackage{amsxtra}
\usepackage{mathrsfs}
\usepackage{textgreek}
\usepackage{pgfplots}
\usepackage{xstring} 
\usepackage{enumitem}
\usepackage{multirow}
\usepackage{makecell}
\usepackage{color,soul}
\usepackage{amsthm}
\usepackage{balance}

\newcommand\sys{NuPS}
\newcommand\rectM{$10\text{m}\times1\text{m}$}

\newcommand\prep{\texttt{PrepareSample}}
\newcommand\pullsamp{\texttt{PullSample}}
\newcommand\conform{\texttt{CONFORM}}
\newcommand\bounded[1]{\texttt{BOUNDED#1}}
\newcommand\longterm{\texttt{LONG-TERM}}
\newcommand\nonconform{\texttt{NON-CONFORM}}
\newcommand\bad[1]{\textcolor{darkred}{#1}}

\newcommand\figurespace{}

\newcommand{\sset}[1]{\left\{\,#1\,\right\}} 

\definecolor{primary}{HTML}{33a02c}
\definecolor{secondary}{HTML}{1f78b4}
\definecolor{third}{HTML}{ff7f00}
\definecolor{bg}{HTML}{999999}
\definecolor{darkred}{HTML}{DC143C}

\copyrightyear{2022} 
\acmYear{2022} 
\setcopyright{rightsretained} 

\acmConference[SIGMOD '22]{Proceedings of the 2022 International Conference on Management of Data}{June 12--17, 2022}{Philadelphia, PA, USA}
\acmBooktitle{Proceedings of the 2022 International Conference on Management of Data (SIGMOD '22), June 12--17, 2022, Philadelphia, PA, USA}
\acmDOI{10.1145/3514221.3517860}
\acmISBN{978-1-4503-9249-5/22/06}

\usepackage{etoolbox}
\makeatletter
\patchcmd{\maketitle}{\@copyrightpermission}{
   \begin{minipage}{0.3\columnwidth}
     \href{http://creativecommons.org/licenses/by/4.0/}{\includegraphics[width=0.95\textwidth]{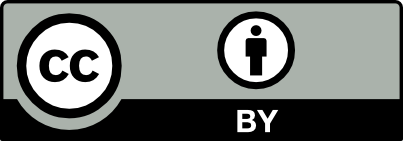}}
   \end{minipage}\hfill
   \begin{minipage}{0.7\columnwidth}
     \href{http://creativecommons.org/licenses/by/4.0/}{This work is licensed under a Creative Commons Attribution International 4.0 License.}
   \end{minipage}
  
   \vspace{5pt}
}{}{}

\makeatother

\begin{document}
\fancyhead{}
\title{NuPS: A Parameter Server for Machine Learning with Non-Uniform Parameter Access}


\author{Alexander Renz-Wieland}
\affiliation{%
    \institution{Technische Universität Berlin}
  }

\author{Rainer Gemulla}
\affiliation{%
    \institution{Universität Mannheim}
  }

\author{Zoi Kaoudi}

\author{Volker Markl}
\affiliation{%
  \institution{Technische Universität Berlin}
}
\affiliation{%
  \institution{BIFOLD}
}


\begin{abstract}
  Parameter servers (PSs) facilitate the implementation of distributed training
  for large machine learning tasks.
  In this paper, we argue that existing PSs are inefficient for tasks that
  exhibit non-uniform parameter access; their performance may even fall behind
  that of single node baselines.
  We identify two major sources of such non-uniform access: skew and sampling.
  Existing PSs are ill-suited for managing skew because they uniformly apply the
  same parameter management technique to all parameters. They are inefficient
  for sampling because the PS is oblivious to the associated randomized accesses
  and cannot exploit locality.
  To overcome these performance limitations, we introduce \sys{}, a novel PS
  architecture that (i) integrates multiple management techniques and employs a
  suitable technique for each parameter and (ii) supports sampling directly via
  suitable sampling primitives and sampling schemes that allow for a controlled
  quality--efficiency trade-off.
  In our experimental study, \sys{} outperformed existing PSs by up to one order
  of magnitude and provided up to linear scalability across multiple machine
  learning tasks.
\end{abstract}



\maketitle

\input{01-introduction}
\input{02-non-uniform-access}
\input{03-skew}
\input{04-sampling}
\input{05-experiments}
\input{06-conclusion}

\begin{acks}
  This work was supported by the German Ministry of Education and Research in
  the Software Campus (01IS17052) and BIFOLD (01IS18037A) programs and by the
  German Research Foundation in the moreEVS (410830482) program.
\end{acks}

\newcommand{\showCODEN}[1]{\unskip}
\newcommand{\showISSN}[1]{\unskip}
\newcommand{\showLCCN}[1]{\unskip}
\newcommand{\showISBNx}[1]{\unskip}
\newcommand{\showISBNxiii}[1]{\unskip}
\newcommand{\showDOI}[1]{\unskip}
\newcommand{\shownote}[1]{\unskip}
\newcommand{\showURL}[1]{\unskip}
\bibliographystyle{ACM-Reference-Format}
\balance
\bibliography{main}


\end{document}
\endinput

%% file: 01-introduction.tex
\section{Introduction}

To keep up with increasing dataset sizes and model complexity, distributed
training has become a necessity for large machine learning (ML) tasks.
Distributed training enables (i) scaling to models and datasets that exceed the
memory of a single machine by distributing them to the nodes of a compute cluster
and (ii) faster training by performing distributed compute. Usually, each node
accesses only its local part of the training data, but requires global read and
write access to all model parameters. Parameter management is thus a key concern
in distributed ML. \emph{Parameter servers} (PS) ease distributed parameter
management by providing primitives for reading and writing parameters, while
transparently handling partitioning and synchronization across
nodes~\cite{smola10, ahmed12, distbelief, ssp, pslite}. Many ML system stacks
employ a PS as a core component, e.g., TensorFlow~\cite{tensorflow},
MXNet~\cite{mxnet}, PyTorch BigGraph~\cite{biggraph}, STRADS~\cite{strads},
STRADS-AP~\cite{stradsap}, or Project Adam~\cite{adam}, and there are many
standalone PSs, e.g., Petuum~\cite{ssp}, PS-Lite~\cite{pslite},
Angel~\cite{angel}, FlexPS~\cite{flexps}, Glint~\cite{glint}, PS2~\cite{ps2},
Lapse~\cite{lapse}, and BytePS~\cite{byteps}.

\input{fig_firstpage}

As cluster nodes access parameters over the network, distributed training
induces communication overhead. For some ML tasks, this overhead causes the
performance of distributed implementations to even fall behind that of single
node baselines~\cite{lapse}; Figure~\ref{fig:first-page} depicts this
exemplarily for a large knowledge graph embeddings task. We observe that a key
cause for such poor performance can be \emph{non-uniform parameter access} and
focus on ML tasks where this is the case. We identify two main sources of
non-uniformity: \emph{skew} and \emph{sampling}. First, in a workload that
exhibits skew, a (typically small) subset of parameters is accessed
frequently (e.g., up to \num{100000} times per second), whereas a large part of
the parameters is accessed rarely (e.g., only once every
minute)~\cite{mf-power-law-samples, kg-characteristics, powergraph,
  powerlaw-internet, zipf-english-texts, powerlaw-empirical-data}. The main
reason for skew is that real-world datasets often have skewed frequency
distributions (e.g., graphs~\cite{kg-characteristics, powergraph,
  powerlaw-internet}, texts~\cite{zipf-english-texts}, and
others~\cite{powerlaw-empirical-data, mf-power-law-samples}), and many ML models
associate specific parameters with specific data items (e.g., with the tokens in
a text document or with the vertices of a graph)~\cite{word2vec, rescal,
  mf-recsys}). The second source of non-uniformity is \emph{sampling}: for a
subset of parameter accesses, random sampling (rather than training data)
determines which parameters are read and written~\cite{word2vec, kgetraining,
  analogy, recsys-implicit-feedback, Bamler2020Extreme, triplet-loss-2010,
  triplet-loss-2015}. One common reason for this access pattern is negative
sampling~\cite{word2vec, Bamler2020Extreme, kgetraining, node2vec}, which, for example, is
used to reduce the cost of many-class classification tasks or to mitigate an
absence of negative training data (e.g., in recommender systems with
only positive feedback or in knowledge graphs that contain only positive edges).

In this paper, we explore how to extend the scope of PSs to ML tasks that
exhibit such non-uniform parameter access. To this end, we present \sys{}, a
novel \ul{n}on-\ul{u}niform \ul{PS} architecture. Figure~\ref{fig:architecture}
depicts an overview of this architecture. \sys{} overcomes two key performance
limitations of existing PSs.
First, existing PSs are inefficient for managing skew because they employ one
single management technique for all parameters. Using a single technique limits
performance as none of the existing techniques is efficient for all access
patterns. To overcome this limitation, \sys{} introduces \emph{multi-technique
  parameter management}, i.e., it integrates multiple parameter management
techniques and chooses a suitable technique \emph{for each parameter}.
In particular, NuPS integrates both replication~\cite{ssp, essp} and
relocation~\cite{lapse}.

\input{fig_architecture}

Second, existing PSs are inefficient for sampling because common parameter
management techniques are ill-suited for randomly sampled access. To improve
performance, applications can implement specialized \emph{sampling schemes}
manually, outside the PS~\cite{biggraph, dglke, word2vec-hogbatch,
  w2v-dist-negs}, but this limits the efficiency of some schemes, potentially
produces incorrect samples, and causes repeated implementation effort. \sys{}
overcomes this limitation by integrating sampling schemes directly into the PS.
To do so, \sys{} extends the PS API with a sampling primitive that allows
applications to request samples from a specific sampling distribution (rather
than accessing specific parameters directly). \sys{}'s \emph{sampling manager}
transparently chooses one of several sampling schemes to reduce
communication overhead for sampling, according to a \emph{conformity level}.
Conformity levels provide a controlled trade-off between efficiency and sample
quality.

In our experimental evaluation, \sys{} outperformed state-of-the-art
PSs by up to one order of magnitude and provided up to linear scalability across
multiple ML tasks. Figure~\ref{fig:first-page} exemplarily shows its performance
for the task of training knowledge graph embeddings.

In summary, our contributions are as follows: (i) we evaluate the suitability of
existing PSs under skew (Section~\ref{sec:skew:sota}), (ii) we propose
multi-technique parameter management to handle skew efficiently
(Section~\ref{sec:skew:nups}), (iii) we develop a hierarchy of
conformity levels (Section~\ref{sec:sampling:scl}) and analyze properties of
common sampling schemes (Section~\ref{sec:sampling:techniques}), (iv) we argue
for and propose a PS API extension for sampling (Section~\ref{sec:sampling:api})
and present how \sys{} implements several schemes behind this API
(Section~\ref{sec:sampling:impl}), and (v) we experimentally investigate how
these changes affect PS performance (Section~\ref{sec:experiments}).


%% file: fig_firstpage.tex
\begin{figure}
  \centering
  \includegraphics[page=1,width=1\columnwidth]{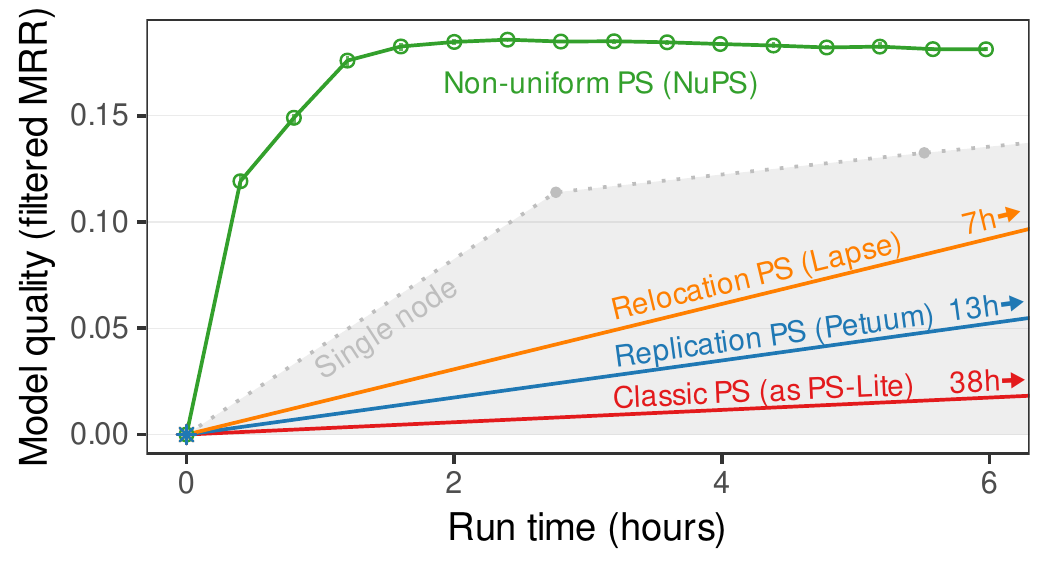}
  \caption{Parameter server (PS) performance for training large knowledge graph
    embeddings (ComplEx~\cite{complex}, dimension~500 on Wikidata5m data) on
    an 8-node cluster (8~worker threads per node). The performance of
    state-of-the-art PSs falls behind that of a single node (8 worker threads)
    due to communication overhead. \sys{} improves performance by up
    to one order of magnitude. Details in Section~\ref{sec:experiments:setup}.}
  \label{fig:first-page}
  \figurespace{}
\end{figure}


%% file: fig_architecture.tex
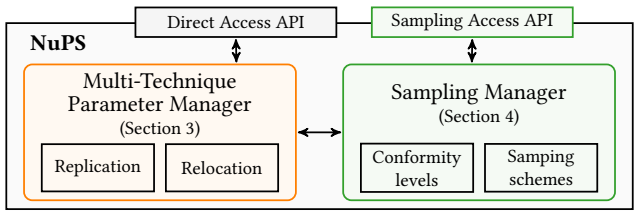
\begin{figure}
  \centering
  \resizebox{1.0\columnwidth}{!}{
    \begin{tikzpicture}[line width = 0.3mm,
      comp/.style={draw, fill=bg!10, rounded corners},
      part/.style={draw, align=center, minimum height=0.5cm, minimum width=3.4cm, fill=bg!10},
      mgmt/.style={draw, align=center, minimum height=0.8cm, minimum width=1.9cm, yshift=-0.6cm},
      left/.style={xshift=-1.05cm},
      right/.style={xshift=1.05cm},
      samt/.style={draw, align=center, minimum height=0.7cm, minimum width=4.0cm, fill=bg!10, yshift=-0.8cm},
      connection/.style={draw, <->,>=stealth', shorten <=0.5pt, shorten >=0.5pt},
      sampling/.style={draw=primary!50},
      ]

      \node (sys) [draw, fill=bg!5, minimum height = 3.15cm, minimum width = 10.6cm, anchor=north west] at (-0.3, 3.4) {};
      \node [anchor=north west, xshift=0.3cm, yshift=-0.10cm] at (sys.north west) {\Large \textbf{\sys{}}};
      
      \node (api) [comp, minimum height = 1.4cm, minimum width = 8cm, anchor=west, draw=none, fill=none] at (2, 3.4) {};
      \node (pullpush) [part, xshift=1.6cm] at (api.west) {Direct Access API};
      \node (sampling) [part, draw=primary, fill=primary!5, xshift=4.0cm] at (pullpush) {Sampling Access API};
      
      \node (pmgmt) [comp, draw=third, fill=third!5, minimum height = 2.3cm, minimum width = 4.6cm, anchor=north west] at (0, 2.7) {};
      \node [yshift=-0.7cm, align=center] at (pmgmt.north) {\Large Multi-Technique\\ \Large Parameter Manager\\(Section~\ref{sec:skew})};
      \node (replication) [mgmt, left] at (pmgmt) {Replication};
      \node (relocation) [mgmt, right] at (pmgmt) {Relocation};

      \node (smgr) [comp, draw=primary, fill=primary!5, minimum height = 2.3cm, minimum width = 4.6cm, anchor=north west, align=center] at (5.4, 2.7) {};
      \node [yshift=-0.7cm, align=center] at (smgr.north) {\Large Sampling Manager\\(Section~\ref{sec:sampling})};
      \node (reloc) [mgmt, left] at (smgr) {Conformity\\levels};
      \node (reuse) [mgmt, right] at (smgr) {Samping\\schemes};

      \draw [connection] (pullpush.south) to (pmgmt.42);
      \draw [connection] (sampling.south) to (smgr.96);
      \draw [connection] (smgr.west) to (pmgmt.east);

    \end{tikzpicture}
  }
  \caption{\sys{} architecture. \sys{} differs from existing PSs in two main
    ways: it introduces (i) \protect\setulcolor{third} \ul{multi-technique
      parameter management} \protect\setulcolor{primary} to handle skew and (ii)
    a \ul{sampling manager and API} to handle sampling. }
  \label{fig:architecture}
  \figurespace{}
  \vspace{-0.2cm} 
\end{figure}


%% file: 02-non-uniform-access.tex
\section{Non-Uniform Parameter Access}

We study ML tasks that exhibit non-uniform parameter access. We identify two
main sources of non-uniformity: skew (Section~\ref{sec:bg:skew}) and sampling
(Section~\ref{sec:bg:sampling}).

\subsection{Skew}
\label{sec:bg:skew}

A workload exhibits skew non-uniformity when some parts of the model are accessed (much)
more frequently than others. The main reason for this is that many real-world
datasets have skewed frequency distributions~\cite{mf-power-law-samples,
  kg-characteristics, popularity-in-kge, powergraph, powerlaw-internet, zipf-english-texts,
  powerlaw-empirical-data}. For example, heavy skew is common in text corpora,
because word frequencies are skewed~\cite{zipf-english-texts}, and in graph
data, because in- and out-degree distributions are
skewed~\cite{kg-characteristics, popularity-in-kge, powergraph, powerlaw-internet}. As many ML
models associate specific parameters with specific data items (e.g, with words
in a text or with the nodes of a graph)~\cite{word2vec, rescal, mf-recsys,
  node2vec}, access to the parameters is heavily skewed, too: a small subset of
\emph{hot spot parameters} is accessed frequently, whereas the majority of
parameters is accessed rarely. In
the following, we will refer to the parameters that are not hot spots as
\emph{long tail parameters}.

We have measured the extent of skew for two real-world ML tasks: training
knowledge graph embeddings and training word vectors. The left hand sides of
Figures~\ref{fig:ad:kge} and \ref{fig:ad:wv} show the number of reads per
parameter over one epoch of these tasks, respectively.  Access is heavily skewed: in the knowledge
graph embeddings task, 18\% of 12.9 trillion total reads go to only 0.02\% of
4.8 billion parameters. In the word vectors task, 45\% of 9 trillion total reads
go to 0.17\% of 1.9 billion parameters. Details on the tasks and datasets can be
found in Section~\ref{sec:experiments:setup}.

\input{fig_access_distributions}

Note that skew is not always present in distributed training.
For example, there is no skew in convolutional neural networks for
image recognition~\cite{cnn} because model access is dense, i.e., every update
step writes to all parameters. In contrast, in common neural network models for
natural language processing~\cite{elmo, bert, ulmfit}, access is partially dense,
and partially sparse and skewed: access to the first (embedding) layer and sometimes the
last (classification) layer is based on word or token frequency (and thus
sparse and skewed), and access to other layers is dense. The share of parameters with
frequency-based access depends on the model architecture, but can be high, e.g.,
around 90\% in ELMo~\cite{elmo}. In this paper, we investigate skew in shallow
models, but conjecture that a non-uniform PS can also be beneficial for deeper
models with partially skewed access.

\subsection{Sampling}
\label{sec:bg:sampling}

A workload exhibits sampling non-uniformity when, for a subset of parameter
accesses, random sampling determines which parameters are read and
written~\cite{word2vec, kgetraining, analogy, recsys-implicit-feedback,
  disentangling-sampling, Bamler2020Extreme, triplet-loss-2010,
  triplet-loss-2015}. I.e., the application randomly draws a parameter key
  from an application-specific sampling distribution over (all or a subset of)
  parameter keys. It then accesses the drawn parameter for training. We refer
to such access as \emph{sampling access}. In contrast, in \emph{direct access},
the training data determines which parameters are accessed. Sampling access is
common in many-class classification tasks, e.g., extreme
classification~\cite{Bamler2020Extreme}, natural language
processing~\cite{word2vec}, knowledge graph embeddings~\cite{kgetraining,
  analogy}, graph representations~\cite{node2vec, negative-sampling-graphs},
recommender systems~\cite{recsys-implicit-feedback}, and when triplet loss is
used~\cite{triplet-loss-2010, triplet-loss-2015}.

For example, knowledge graph embeddings and word vectors training tasks often
use \emph{negative sampling} to enable efficient training~\cite{word2vec, kgetraining,
  disentangling-sampling}. For each (positive) data
point, a set of \emph{negative samples} is drawn from a distribution. Each
negative sample corresponds to a data item (e.g., a word) or a class. The
corresponding parameters are subsequently accessed for training.
For instance, the example knowledge graph embeddings task draws negative
samples from a uniform distribution over all
entities~\cite{kgetraining,analogy}. The right hand side of Figure~\ref{fig:ad:kge}
shows the frequency distributions of direct and sampling accesses separately for
this task. In our implementation (based on~\cite{analogy}) and with 200 negative
samples for each subject--relation--object triple (100 negative samples for the
subject and another 100 for the object), sampling accesses make up 31\%
of all accesses.
%
In the word vectors task, negative samples correspond to words and the sampling
distribution resembles the word frequencies in the training
data~\cite{word2vec}, see Figure~\ref{fig:ad:wv}. In the plot for direct access,
parameters that belong to the output layer of the task's neural network are
visually distinct from the other parameters. The reason for this is that the
task draws samples only from the output layer, and parameters in the plot are
sorted by \emph{total} access frequency. In our implementation (based
on~\cite{word2vec}) and with 3 negative samples for each word--word pair,
sampling accesses make up 56\% of all parameter accesses in this task.


%% file: fig_access_distributions.tex
\tikzset{	
	col_small_arrows/.style={->,>=stealth', line width=0.26mm, shorten <=2pt, shorten >=2pt}
}

\newcommand\accessdist[1]{
  \begin{tikzpicture}
    \node (all) [anchor=east,inner sep=0pt] at (0,0) {\includegraphics[width=0.5\columnwidth]{./plots/access_distr.#1.all.png}};

    \node (key) [anchor=south west, inner sep=0pt] at (1,-0.03) {\includegraphics[width=0.40\columnwidth]{./plots/access_distr.#1.key.png}};
    \node (sampling) [anchor=north west, inner sep=0pt] at (1,0.09) {\includegraphics[width=0.40\columnwidth]{./plots/access_distr.#1.sampling.png}};

    \draw [col_small_arrows] (all.10) to (key.180);
    \draw [col_small_arrows] (all.-0) to (sampling.175);
  \end{tikzpicture}
}

\begin{figure}

  \begin{subfigure}[b]{0.9\columnwidth}
  \accessdist{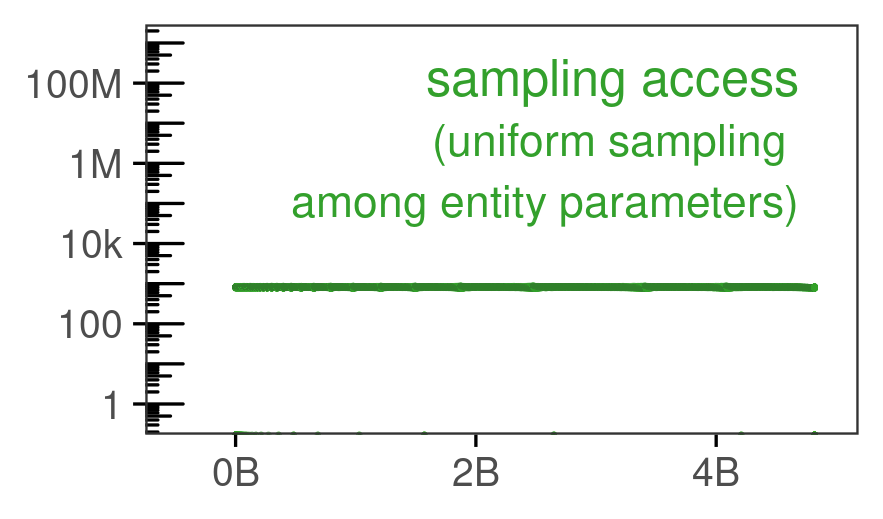}
  \vspace*{-0.3cm}
  \caption{Knowledge graph embeddings}
  \label{fig:ad:kge}
  \end{subfigure}

  \vspace{0.3cm}
  \begin{subfigure}[b]{0.9\columnwidth}
    \accessdist{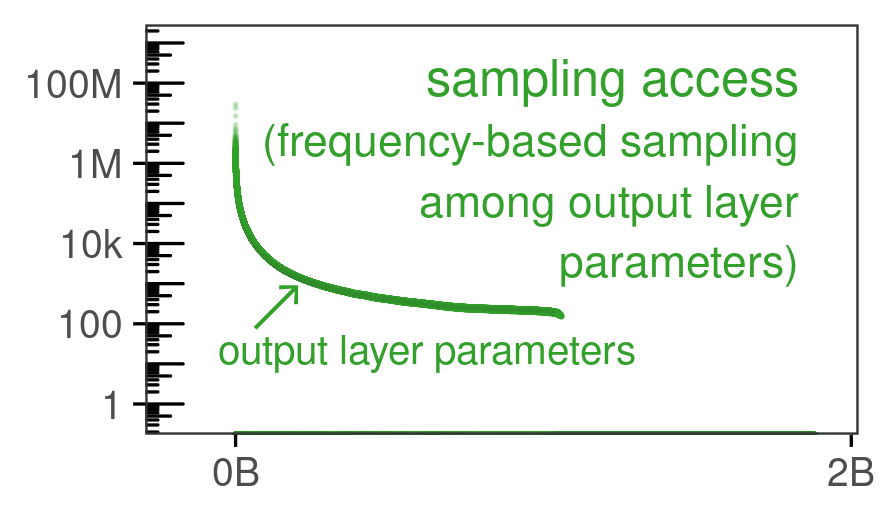}
    \vspace*{-0.3cm}
    \caption{Word vectors}
    \label{fig:ad:wv}
  \end{subfigure}
  \captionsetup{skip=6pt} 
  \caption{Number of accesses per parameter in one epoch. Parameters
    are sorted by decreasing total number of accesses. See
    Section~\ref{sec:experiments:setup} for details on tasks and
    experimental setup.}
  \label{fig:ad}
  \figurespace{}
  \vspace{-0.2cm} 
\end{figure}


%% file: 03-skew.tex
\section{Multi-Technique Parameter Management}
\label{sec:skew}

In this section, we analyze the suitability of existing PSs for ML tasks with
skewed parameter access (Section~\ref{sec:skew:sota}) and argue that existing
PSs are inefficient for managing skew because they employ one single management
technique for all parameters. Based on this analysis, we propose multi-technique
parameter management and discuss \sys{}'s implementation
(Section~\ref{sec:skew:nups}).

\subsection{Analysis of Common Parameter Management Techniques}
\label{sec:skew:sota}

PSs~\cite{smola10, ahmed12, distbelief, ssp, pslite, glint, ps2, angel, flexps,
  lapse, byteps} partition the model parameters across a set of \emph{servers}.
The PS provides \texttt{pull} and \texttt{push} primitives for global reads and
writes to model parameters, respectively. In the data-parallel setting, the
training data are partitioned to a set of \emph{workers}. During training, each
worker processes its local part of the training data (often multiple times) and
continuously reads and updates model parameters. To coordinate parameter
accesses across workers, each parameter is assigned a unique \emph{key}. Many
PSs physically co-locate the (logically distinct) servers and workers on the
same nodes for efficiency, either in multiple processes per node~\cite{pslite,
  glint, angel} or within one process~\cite{ssp, flexps, lapse}.

Several techniques have been proposed for managing parameters among the cluster
nodes in a PS. In the following, we discuss common techniques,
briefly introducing each before analyzing its suitability for managing skew.

\subsubsection{Classic PS}

A classic PS allocates parameters to servers statically (e.g., via
range partitioning of the parameter keys) and uses no replication~\cite{smola10,
  ahmed12, pslite}. Thus precisely one server holds the current value of a
parameter, and this server is used for all pull and push operations on this
parameter. Classic PSs typically guarantee sequential consistency for operations
on the same key~\cite{lapse}.

\textbf{Analysis: The performance of a classic PS is limited for both hot spots
  and long tail parameters.} The reason for this is that every parameter access uses
the network: it incurs network latency for two messages (to and from the
responsible server) and the parameter value is sent over the network once (from
the server to the worker in a pull operation, in the other direction for a push
operation). This network overhead is incurred for all parameters, i.e., hot spot
and long tail ones. For hot spots, the overhead is incurred many
times for a few parameters. In the long tail, the overhead is incurred a few
times for each of many parameters.

\subsubsection{Replication PS}

A replication PS replicates parameters and tolerates some amount of staleness in
the replicas~\cite{ssp, flexps, angel, essp, iterstore}. Replication PSs provide
weaker consistency guarantees, such as \emph{bounded staleness}, and require
applications to explicitly control staleness via special primitives (e.g., an
``advance the clock'' operation). There are two main protocols for creating and
refreshing replicas in general-purpose PSs: \emph{SSP}~\cite{ssp} creates a
replica when a parameter is accessed and uses this replica until the staleness
bound is reached (at which point the replica is terminated).
\emph{ESSP}~\cite{essp} also creates a replica when a parameter is (first)
accessed, but then maintains this replica throughout the entire training task
(by repeatedly propagating updates). In both SSP and ESSP, nodes accumulate
replica updates locally and propagate them to the responsible server at the
subsequent ``advance the clock'' invocation. A subset of replication PSs
specifically target deep learning workloads in which each node holds replicas of
\emph{all} parameters and replicas are updated synchronously after each step of
mini-batch stochastic gradient descent~\cite{poseidon,
  ps-priority-parameter-prop, tictac, byteps}. In contrast to \sys{}, these PSs
focus on workloads in which (i) the model size does not exceed the memory
capacity of a single node and (ii) synchronous replica updates are not
prohibitively slow w.r.t. to computational cost; some further apply only to
GPU-based training~\cite{byteps}.

\textbf{Analysis: A replication PS is efficient for hot spots, but its benefit for
  the long tail is limited.} Replication reduces network overhead (compared to a
classic PS) if a replicated parameter value is used more than once and multiple
updates can be sent to the PS in aggregated form. Replication further reduces
access latency if a parameter value (within the acceptable staleness bound) is
already locally available when a read operation is issued. Both is typically the
case for hot spot parameters, even within relatively tight staleness bounds
(because hot spot parameters are accessed frequently at each node). In contrast,
long tail parameters are accessed infrequently. So it is unlikely that a long
tail parameter is accessed more than once within reasonable staleness bounds
(large staleness bounds commonly deteriorate model convergence~\cite{ssp}). For
the same reason, SSP (which creates replicas on demand) does not reduce access
latency for long tail parameters, because replicas are mostly ``cold''.
With its eager replica maintenance, ESSP ensures that replicas are
always ``warm'' (after the first access to a parameter), but at the
cost of significant over-communication: ESSP constantly updates all replicas,
although replicas for long tail parameters are accessed rarely.

\subsubsection{Relocation PS}

A relocation PS asynchronously re-allocates parameters among nodes during
run time so that access operations can be processed locally, without network
communication~\cite{lapse}. Relocation PSs require applications to control
allocation via special primitives (e.g., a ``localize'' operation). As classic
PSs, relocation PSs can provide per-key sequential consistency~\cite{lapse}.

\textbf{Analysis: A relocation PS is efficient for long tail parameters, but has
  limited benefit for hot spots.}
Relocation eliminates access latency if there is sufficient time to relocate a
parameter between accesses at different nodes. It further reduces network
overhead (compared to classic) if a parameter is accessed more than once between
two relocations (which is common, most ML tasks at least read and write a
parameter): a relocation takes three messages in Lapse~\cite{lapse}
(including the parameter value once), whereas each remote access in a classic PS
sends two messages (including the parameter value once). There is typically
sufficient time for relocating long tail parameters between accesses by
different nodes, as they are accessed infrequently. Hot spot
parameters, however, are frequently accessed at multiple nodes concurrently~\cite{lapse-demo}.
Thus, there is not sufficient time for relocations between accesses, such that
access latency is not eliminated. Further, a relocation PS incurs higher network
overhead than a classic PS if a parameter is relocated so frequently that only
one operation is processed locally.

\subsubsection{Summary}

Individual management techniques are efficient for either hot spot
\emph{or} long tail parameters (or neither of the two), but none is efficient
for both. Consequently, managing all parameters with the same technique limits
the performance of PSs for ML tasks with skewed parameter access. 

\subsection{Parameter Management in NuPS}
\label{sec:skew:nups}

From the above discussion, it follows naturally to explore whether combining multiple
management techniques is beneficial for PS performance. The idea of combining
multiple management techniques has been studied in other (non-PS) distributed
data management systems, such as general-purpose distributed
databases~\cite{file-assignment-problem-survey, db-dynamic-replication,
  managing-replicted-distr-db, analysis-replication-distr-db} and distributed
graph processing systems~\cite{graphlab-distributed}. These systems combine
static allocation with replication, but do not consider relocation. To the best
of our knowledge, integrating multiple management techniques in PSs has not been
explored before.

\sys{} integrates two management techniques: replication and relocation. First,
to manage hot spot parameters efficiently, \sys{} integrates a lightweight
variant of eager replication~\cite{essp}. \sys{} eagerly creates replicas for
hot spot keys on all nodes and provides time-based staleness bounds. Basing the
staleness bound on time rather than clocks alleviates the need for adding
``advance the clock'' operations to application code, but potentially
complicates the analysis of convergence properties. We discuss these
implications below. Second, to manage long tail
parameters efficiently, \sys{} integrates relocation. As Lapse~\cite{lapse}, \sys{}
asynchronously relocates these parameters before they are accessed. This
guarantees per-key sequential consistency for long tail parameters. \sys{}
picks a technique for each key based on the key's access pattern: if the key is accessed
frequently, \sys{} replicates the key; if there
are few accesses, \sys{} employs relocation (see
Section~\ref{sec:experiments:setup}). The choice of management technique is
transparent to the application, i.e., the application accesses all parameters in the same way,
via the \texttt{push} and \texttt{pull} primitives. Our experimental evaluation
shows that the combination of
replication and relocation can be highly beneficial. Integrating other
techniques (e.g., highly tailored ones) may further improve performance, but is
beyond the scope of this paper. \sys{} does not integrate the classic technique
as it is dominated by replication for hot spots and by relocation for the long tail.

\input{fig_local_remote}

For efficiency, \sys{} co-locates workers and servers in one process per node,
and accesses replicas and locally allocated parameters via shared memory.
Figure~\ref{fig:parameter-management} depicts an overview. To
access a key, a worker checks whether this key is managed by replication or
relocation. If the key is managed by replication, the worker accesses the key
via shared memory, without network communication. If the key is managed by
relocation, the worker checks whether the key is currently allocated locally. If
so, it accesses the key via shared memory. Otherwise, the worker accesses the
parameter remotely, using the message protocol proposed in Lapse~\cite{lapse}: a
request to the node that knows where the parameter is currently allocated, which
then forwards the request to this node, which
in turn processes the request and sends a response to the worker. 

\sys{} is designed to minimize the run time overhead of providing multiple
management techniques. To do so, \sys{} integrates the check for the management
technique and the check for local allocation into one latch acquisition (i.e., a
lock held for the duration of the API call). Further, \sys{} can be reduced to a
single-technique PS with no measurable run time overhead for providing more than
one management technique: If replication is not used for any key, the replica
synchronization background thread exits immediately, without sending any
messages. If relocation is not used for any key, no messages are sent for
relocation.

\sys{} bases its staleness bounds on time rather than logical clocks
  because this makes the PS easier to use: time-based bounds alleviate the need
  for adding ``advance the clock'' operations to application code and for timing
  them appropriately. \sys{} synchronizes the replicas periodically, using
  sparse all-reduce operations (i.e., only updated parameters are
  exchanged~\cite{mpi-neutral-element-elimination}). The synchronization is run
  by a background thread and uses the recursive doubling algorithm. However,
  time-based bounds potentially complicate the analysis of convergence
  properties. If only a bounded number of SGD steps can occur within one
  synchronization round, bounded staleness holds (as for clock-based staleness
  bounds) and the corresponding analysis carries over~\cite{ssp}. However, if
  the number of SGD steps within one synchronization round cannot be bounded,
  convergence analyses for asynchronous SGD apply~\cite{async-sgd-nonconvex,
    taming-async-sgd}. In our experiments, the effect of time-based bounds on
  performance was minimal because we used replication only for a small number of
  parameters and synchronized replicas frequently (see
  Section~\ref{sec:exp:techniques} and Section~\ref{sec:exp:so}).


%% file: fig_local_remote.tex
\definecolor{neutralbg}{HTML}{eeeeee}

\tikzset{	
	col_object/.style={minimum width=5.10cm, minimum height = 2.70cm,draw=black, rounded corners, fill=neutralbg, align=center}, 
	col_arrow_style/.style={<->,>=stealth',line width=0.2mm, color=black, shorten <=3pt, shorten >=3pt}, 
	col_rectangles/.style={draw=black,fill=secondary!30, rounded
    corners=0.7mm,minimum height=5.0mm,minimum width=1.55cm, anchor=north west,
    yshift=-0.5mm, xshift=0mm, align=center, text centered},
	col_small_arrows/.style={<->,>=stealth', line width=0.26mm, shorten <=2pt, shorten >=2pt},
  store/.style={minimum width=1.8cm, anchor=north east, align=center, font=\large}
}

\def \colfontsize {\large}

\newcommand\showtext[1]{#1}

\newcommand{\worker}[6]{
	\node [col_object,scale=#6](nd#1) {};
	\node [col_object,scale=#6](nd#1) [label=below:{\colfontsize #2}] {};

  \path [fill=third!35, anchor=east, draw,scale=#6]
  ([yshift=3.0mm, xshift=-14.0mm]nd#1.east) --
  ++(1.3,0) {[rounded corners=0.7mm] --
    ++(0,0.55) --
    ++(-1.3,0)} --
  cycle
  {};
  \path[fill=third!50,draw,scale=#6]
  ([yshift=3.0mm, xshift=-14.0mm]nd#1.east) --
  ++(1.3,0) {[rounded corners=0.7mm] --
    ++(0,-#4) --
    ++(-1.3,0)} --
  cycle
  {};

	\node (nd#1pa) [store, yshift=-1.0mm, xshift=-1.5mm,scale=#6] at (nd#1.north east) {\showtext{Replicated\\[-3pt]parameters}};
  \node (nd#1reloc) [store, minimum height=#3, yshift=0.5mm,text width=1.6cm,scale=#6] at (nd#1pa.south east) {\showtext{Local\\#5relocated\\#5parameters}};

	\node (nd#1a) [scale=#6,col_rectangles, anchor=north west, xshift=1.5mm, yshift=-1.0mm] at (nd#1.north west) {\showtext{\colfontsize{worker 1}}};

	\node (nd#1b) [scale=#6, col_rectangles] at (nd#1a.south west) {\showtext{\colfontsize{worker 2}}};

	\node (nd#1c) [scale=#6, col_rectangles,draw=none,fill=none,minimum height=0mm,yshift=+0.6mm] at (nd#1b.south west) {...};

	\node (nd#1d) [scale=#6, col_rectangles,yshift=+0.6mm] at (nd#1c.south west) {\showtext{\colfontsize{worker n}}};

	\node (nd#1e) [scale=#6, col_rectangles,fill=secondary!10] at (nd#1d.south west) {\showtext{\colfontsize{server}}};

	\draw [scale=#6,col_small_arrows, shorten <=0.1cm, shorten >=0.1cm]
  (nd#1c.east) to  ([xshift=1cm]nd#1c.east) node[midway, align=center, font=\large,
  anchor=center, xshift=-0.15cm, yshift=0.5cm,scale=#6] {\showtext{shared\\memory}};

}

\begin{figure}
  \centering

  \begin{tikzpicture}[ 
   	line width = 0.2mm,
    scale=0.70,
    every node/.style={scale=0.70}
    ]

    \matrix[row sep=-2mm,column sep=-3.5mm]  {
      & \worker{1}{\sys{} process at node 1}{1.2cm}{1.0}{}{1.0}; &  \\
      \renewcommand\showtext[1]{}
      \worker{2}{\sys{} process at node 2}{0.8cm}{0.8}{[-3pt]}{0.6}; & & \renewcommand\showtext[1]{}\worker{3}{\sys{} process at node 3}{1.7cm}{1.16}{}{0.6};
      \\};        

    \draw [col_arrow_style](nd2.east) to [bend right=10] node[above] {} (nd3.west);
    \draw [col_arrow_style](nd3.north) to [bend right=30] node[below,xshift=8mm,yshift=13mm,align=center,font=\large] {remote access,\\replica sync.,\\relocation} (nd1.east);
    \draw [col_arrow_style](nd1.west) to [bend right=30] node[below] {} (nd2.north);
    
  \end{tikzpicture}

  \captionsetup{skip=6pt} 
  \caption{Parameter management in \sys{}. \sys{} replicates hot spots
    and relocates long tail parameters. It accesses replicated and current
    local parameters via shared memory.}
  \label{fig:parameter-management}
  \vspace{-0.4cm}
\end{figure}
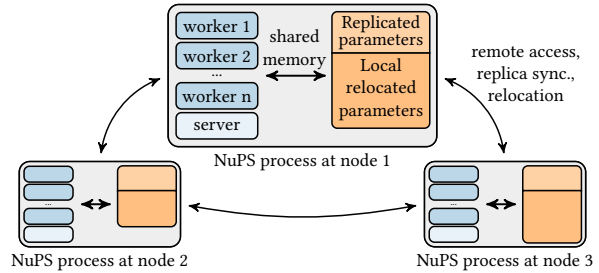


%% file: 04-sampling.tex
\section{Sampling Management}
\label{sec:sampling}

Existing PSs provide no support for sampling. This means that applications
manually sample keys and then access the corresponding parameters via direct
access, which leads to significant communication overhead. To reduce this
overhead, many applications implement a variety of sampling
schemes~\cite{biggraph, dglke, word2vec-hogbatch, w2v-dist-negs}. The key idea
of such sampling schemes is that slightly (or sometimes rather significantly)
deviating from the ideal of independent sampling from the desired target
distribution might have only little or no effect on model quality, but can
reduce communication overhead substantially (and consequently speed up model
training). The lack of sampling support in PSs forces applications to implement
such schemes in application code, outside the PS. This leads to repeated
implementation effort and potentially produces incorrect samples. Further, this
precludes schemes that require tight integration with parameter management.

In contrast to existing PSs, \sys{} integrates sampling directly into the PS. In
the following, we present the components of this integration. We first
introduce a set of conformity levels that allow for a controlled trade-off
between efficiency and sample quality (Section~\ref{sec:sampling:scl}). We then
analyze conformity and communication overhead of sampling schemes that are
commonly used by applications (Section~\ref{sec:sampling:techniques}). Based on
this analysis, we propose an API extension that enables sampling in PSs
(Section~\ref{sec:sampling:api}) and discuss how \sys{} implements several
sampling schemes, within this API (Section~\ref{sec:sampling:impl}).

\subsection{Sampling Conformity Levels}
\label{sec:sampling:scl}

\newcommand\cX{\mathcal{X}}
\newcommand\cK{\mathcal{K}}
\newcommand\cS{\mathcal{S}}
\newcommand\cL{\mathcal{L}}

 Let $\pi$ be a \emph{target
  distribution} over parameter keys. We assume that $\pi$ is specified by the
application and remains fixed throughout run time.\footnote{This is mainly to
  facilitate analysis; an application may use multiple different sampling
  distributions, each of which can be analyzed separately.} For example, in the
word vectors training task of Section~\ref{sec:bg:sampling}, the target
distribution $\pi$ roughly corresponds to relative word
frequencies~\cite{word2vec}; cf.~Figure~\ref{fig:ad:wv}. When training knowledge
graph embeddings, $\pi$ is often a uniform distribution over all
entities~\cite{kgetraining}; cf.~Figure~\ref{fig:ad:kge}.
Denote by $\cK$ the set of parameter keys and by $\pi_k \geq 0$ the target
probability for key $k\in \cK$, where $\sum_{k=1}^{|K|}\pi_k = 1$. Workers repeatedly
draw one or more samples from the target distribution $\pi$. Denote by $X_{qi}\in
\cK$ a random variable for the $i$-th sample obtained at node
$q$.\footnote{Depending on the implementation, there can be multiple workers on
  each node. We analyze sampling schemes at the node level to simplify
  exposition.} We write $N_q$ for the number of samples drawn at node $q$ during
the complete run time of some application. Set $\cX_q=\{X_{q1},\ldots,X_{qN_q}\}$ and
$\cX=\bigcup_q \cX_q$.

We propose a hierarchy of four \emph{sampling conformity levels} to control the
trade-off between sample quality and efficiency. From the top (L1) to the bottom
(L4) of this hierarchy, sample quality decreases, and potential efficiency
increases:
\setlist[description]{leftmargin=\parindent}
\begin{description}
\item[(L1) \conform{}.] The sampling scheme produces mutually independent samples
  from the target distribution $\pi$. I.e.,
  \[
    p(X_{qi}=k|\cS)=\pi_k
  \]
  for all $q,i,k$ and
  $\cS\subseteq \cX\setminus\sset{X_{qi}}$.
\item[(L2) \bounded{}.] The samples at each node have dependencies on past
  samples, but these dependencies are limited and samples at different nodes are
  independent. In more detail, given a \emph{dependency bound} $B\in\mathbb{N}$,
  it holds
  \[
    p(X_{qi}=k|\cS_q^{-B},\cS_{-q})=\pi_k
  \]
  for all $q,i,k$, where $\cS_{-q}\subseteq
  \cX\setminus\cX_q$ refers to samples at other nodes and $\cS_q^{-B}\subseteq\{X_{q1},\ldots,
  X_{q(i-B-1)}\}$ refers to samples at node $q$ taken from all but the last
  $B$ samples taken so far. Note that first-order inclusion probabilities match
  the target probabilities---i.e., $p(X_{qi}=k)=\pi_k$---even though subsequent
  samples may be dependent. For example, a sampling scheme that internally draws
  independent samples from $\pi$ but uses each sample twice is
  \bounded{} with $B=1$.
\item[(L3) \longterm{}.] The mean first-order inclusion probabilities match the
  target probabilities asymptotically at each node, i.e.,
  \begin{equation}
    \label{eq:longterm}
    \lim_{N_q\to\infty}\frac{1}{N_q}\sum_{i=1}^{N_q} p(X_{qi} = k | X_{q1},\dots,X_{q(i-1)}) = \pi_k
\end{equation}
for all $q,k$. Note that this does \emph{not} imply $p(X_{qi}=k)=\pi_k$. Also,
arbitrary dependencies between samples within one or across multiple nodes are
accepted as long as the asymptotic relative frequencies of the samples match the
target. For example, a sequential sampling scheme that selects a random
  key order for the $|K|$ keys and then draws samples in a round-robin fashion satisfies
  \longterm{} but not \bounded{}: each key is selected equally often in the long
  run, but the knowledge of the first $|K|$ samples allows to uniquely determine
  all future samples, so that no dependency bound can be established.

\item[(L4) \nonconform{}.]
  No guarantees about the sampling probabilities or independence.
\end{description}
The levels are hierarchical in that L1 implies L2, and L2 implies L3.
The first implication follows since we can set $\cS=\cS_q^{-B}\cup\cS_{-q}$ for any
choice of $\cS_q^{-B}$ and $\cS_{-q}$.

\def\qedsymbol{}
\begin{proof}[Proof (L2 implies L3)]
  Starting from some offset $1\le o\le B$, fix some node $q$ and consider the subset
  $\sset{X_{q(aB+o)}}_{a\in\mathbb{N}}$ of every $B$-th sample on node $q$, starting
  from the $o$-th sample. Using the definition of \bounded{}, we obtain
  \[
    \frac1{\lfloor(N_q-o)/B\rfloor}  \sum_{a=1}^{\lfloor(N_q-o)/B\rfloor} \hspace{-0.3cm}p(X_{q(aB+o)} =k| X_{q1},\dots,X_{q(aB+o-B)}) = \pi_k
  \]
  for any choice of $N_q$, i.e., the long-term relative frequencies of every
  $B$-th sample match if we start at offset $o$. Since this holds for every offset
  $o$, we conclude that Eq.~\eqref{eq:longterm} holds and L2 implies L3.
\end{proof}

Note that we defined L3 via Eq.~\eqref{eq:longterm} rather than a simpler
first-order probability condition such as $p(X_{qi}=k) = \pi_k$, because correct
first-order conditions are not sufficient to ensure that a sampling scheme is
useful in practice. For example, a sampling scheme that internally draws one
independent sample $X$ from $\pi$, and then uses solely this sample throughout
(i.e., $X_{qi}=X$ for all $q,i$) satisfies such a condition, but is clearly
unsuitable in practice.

\subsection{Analysis of Common Sampling Schemes}
\label{sec:sampling:techniques}

ML applications employ a variety of sampling schemes. In the following, we
  analyze schemes that are common in distributed training~\cite{biggraph, dglke,
    word2vec-hogbatch, w2v-dist-negs, libkge-dist} w.r.t. their
effect on (i) communication overhead and (ii) sampling quality, i.e., into which
conformity level they fall. Table~\ref{tab:sampling:management-techniques}
provides an overview of the latter.

\input{tab_sampling_management_techniques}

\textbf{Independent sampling. } Ideally, applications draw iid.~samples from
the target distribution and use each sample once. This scheme is \conform{}, but
can lead to significant communication overhead: for each sample, the
corresponding parameter values need to be transferred to the node and, after an
update is computed, updates need to be propagated to other nodes.

\textbf{Sample reuse. } Sample reuse reduces
communication overhead by using each sample multiple
times~\cite{word2vec-hogbatch, dglke, biggraph, libkge}. For example, knowledge
graph embeddings training can use shared sampling, i.e., reuse negative samples for all positive
examples in a mini-batch~\cite{libkge}.  Reusing a sample multiple times
avoids the transfer of parameter values for another, fresh sample: using a sample
$U$ times can reduce the communication overhead by a factor of $U$. We refer to
this factor as the \emph{use frequency} and to a sample reuse scheme that uses
each sample $u$ times as \emph{U=u sample reuse}. Sample reuse does not provide \conform{} since
samples are not independent. However, it can provide \bounded{}. For example, if
each fresh sample is sampled iid.~from $\pi$ and then used exactly $U$ times,
then the scheme is \bounded{} for all $B\ge U$. Moreover, in mini-batch negative sample reuse as
in~\cite{word2vec-hogbatch, biggraph, libkge}, \bounded{} also holds. Here
samples are reused only within one mini-batch of gradient descent so that the
mini-batch size provides a bound on the sample dependency.

\textbf{Local sampling. } In many distributed ML architectures~\cite{ssp, essp,
  flexps, lapse}, at each node, a distinct subset of the model parameters---the
\emph{local partition}---can be accessed without network communication.
\emph{Local sampling} restricts sampling accesses to this local
partition~\cite{dglke}. This scheme eliminates network overhead for sampling
accesses entirely. However, local sampling is \nonconform{} as nodes see only
samples from the local partition. Some implementations
re-partition parameters periodically such that all nodes at least see all
samples over time~\cite{dglke}. Careful
re-partitioning might satisfy Eq.~\eqref{eq:longterm} for certain target
distributions; e.g., if $\pi$ is uniform and parameters are allocated uniformly
and at random. In general, however, local sampling cannot provide \longterm{}.
For example, consider any target distributions in which $\pi_k > {1}/{Q}$ for
some $k$ (with $Q$ being the number of nodes). Local sampling cannot satisfy
Eq.~\eqref{eq:longterm} for such a target since key $k$ is available for sampling
at only one node at a time. This implies that there is at least one node at
which the long-term frequency of $k$ is $\le 1/Q$.

\textbf{Direct-access repurposing. } Another sampling scheme
is to repurpose direct-access parameters, i.e., to use them as negative samples. For example, DGL-KE~\cite{dglke}
generates some of the samples by repurposing parameters that occur as positives
in other data points of an SGD mini-batch. This requires no additional
communication for sampling accesses, as the values for the direct access parameters
are transferred to the node either way. In this scheme, the relative
frequency of a seeing a key in a sample depends on the occurrence frequency of
the key in the training data. As the training data occurrence distribution can
be (and typically is~\cite{libkge, word2vec}) different from the target
distribution, this scheme is \nonconform{}.

\subsection{A Primitive for Sampling}
\label{sec:sampling:api}

It is impossible for PSs to integrate these sampling schemes within the
\texttt{push}/\texttt{pull} API. The main problem is that sampling is done by
application code: to conduct a sampling access, an ML application draws a sample
of keys and accesses them via \texttt{pull} or \texttt{push}. For instance, this
makes it impossible for the PS to restrict sampling to the local partition.
Further, the PS cannot even distinguish between direct access (for which it
\emph{cannot} leverage sampling schemes) and sampling access (for which it
\emph{can} leverage sampling schemes).

To overcome these limitations, we propose to extend the PS API with a sampling
primitive that allows applications to access a sample from a target
distribution, under a specific sampling conformity level. The sampling manager
in \sys{} transparently chooses a sampling scheme that conforms with the chosen
conformity level and applies the scheme for all sampling accesses. We propose
one operation \texttt{dist = register\_distribution(π, L)} to register a
specific sampling distribution $\pi$ under a specific sampling conformity
level~$L$, and a combination of two operations to draw samples:
\vspace{-0.3cm}
\begin{align*}
  \mathtt{handle = }&\mathtt{\ PrepareSample(dist, N)} \\
  \mathtt{keys, values = }&\mathtt{\ PullSample(handle[, n_j])}
\end{align*}
The argument $N$ is the number of desired
samples. \prep{} is intended to return instantaneously (and run preparatory work
in the background), \pullsamp{} blocks if called synchronously. After
\pullsamp{} returns, the corresponding keys are stored in \texttt{keys} and
corresponding values are copied to \texttt{values}. Applications can call
\pullsamp{} once to obtain all $N$ samples at once or multiple times to obtain
the $N$ samples in smaller portions (by passing $n_0, n_1, ... < N$ to multiple
invocations of \pullsamp{} such that $\sum n_j = N$). Such partial pulls give the PS
more flexibility, and, thus, may result in better performance.

\input{fig_prep_pull}

This extension provides sufficient flexibility for implementing a wide range of
sampling schemes, as we describe in the following
Section~\ref{sec:sampling:impl}. The extension derives its flexibility from
three key design choices. First, the extension transfers sampling from the
application to the PS. Second, the extension provides the PS with a hook for
doing preparatory work, such as pre-fetching parameter values, modifying
partitions, or coordinating among nodes. Third, the extension does not force
final decisions (e.g., about the sampled keys) before \pullsamp{} returns.

\subsection{The Sampling Manager in NuPS}
\label{sec:sampling:impl}

The sampling manager is responsible for generating samples and managing the
corresponding parameters. The sampling manager of \sys{} currently supports four
sampling schemes behind the sampling API. Figure~\ref{fig:prep_pull} provides an
overview. Schemes implement \prep{} and \pullsamp{}, and optionally a background
thread. From the four implemented schemes, the sampling manager picks a scheme
that is suitable for the specified conformity level. We now discuss the
schemes in turn.

\textbf{Independent sampling (\conform{}).} In this scheme, \sys{} samples
iid.~from the target distribution and localizes the corresponding parameters in
\prep{} (such that they can be accessed locally when \pullsamp{} is called). In
\pullsamp{}, \sys{} accesses the parameters remotely if they have been relocated
to another node in-between \prep{} and \pullsamp{} (this can happen because
other nodes can independently work on the same parameters). This approach is
\conform{} because each worker samples iid.~from~$\pi$.

\input{tab_datasets}

\textbf{Sample reuse (\bounded{}).} \sys{} implements a sample reuse scheme
that reuses pools of keys. The pooling increases the temporal distance
between the reused samples and thereby increases randomness. For a given pool
size~$G$ and use frequency~$U$, \sys{} repeatedly samples $G$ keys iid.~from
$\pi$ to form a \emph{sample pool} and produces samples by traversing the
sample pool $U$ times, each time in a random order. For example, consider $U=2$
and suppose that the iid.~draws produce keys $k_1$, $k_2$, and $k_3$,
respectively. With $G=1$, we obtain sample sequence $k_1k_1k_2k_2k_3k_3$. With
$G=3$, a sequence such as $k_1k_2k_3k_2k_1k_3$ is possible. The pools are
prepared by a background thread. When the background thread generates a new pool, it
localizes the corresponding parameters. \sys{} localizes the parameters again in
\prep{} if they have been relocated to another node since pool preparation. In
\pullsamp{}, \sys{} accesses the parameters remotely if necessary. This sample
reuse scheme is \bounded{} because samples are drawn iid.~from the
target distribution $\pi$, inter-sample dependency is bounded by $U \cdot G$,
and $U$ is identical for all samples.

The background thread determines automatically when to prepare a new
pool. Adding a new pool takes time and (for good performance) the
localization should be finished when \pullsamp{} is called. This time depends on
the ML task, the used hardware, and the system configuration. To estimate this time,
we use a heuristic.\footnote{Note that while the heuristic may
  affect performance, it does not affect correctness.} In particular, the
background thread keeps track of the duration of previous pool relocations. If
the number of prepared, but unused samples is less than double
of the current estimated relocation time, the preparation of another pool is
triggered. 

\textbf{Sample reuse with postponing (\longterm{}). } \sys{} additionally
implements sample reuse with \emph{sample postponing}. This is identical to the
described sample reuse scheme, but adds sample postponing: if sample $i$
cannot be accessed locally in \pullsamp{}, \sys{} re-localizes the corresponding
parameters, postpones sample $i$ for later use, and uses sample $i+1$ instead. To
achieve \longterm{}, it is crucial that, at some point, samples are used (and
not re-postponed indefinitely).\footnote{If samples could be re-postponed
  indefinitely, some samples may never be used because they are constantly being
  relocated. In such cases, Eq.~\eqref{eq:longterm} would not hold.} Thus,
\sys{} postpones only within the~$N$ samples of one invocation of \prep{} (in
other words, only within the samples of one handle). I.e., when \sys{} finds a
non-local sample (in \pullsamp{}), it moves the sample to the end of the $N$ samples of
this handle. \sys{} postpones each sample maximally once. When it reaches
samples that it has already postponed once (towards the end of the $N$ samples),
it accesses them remotely if necessary. This implementation of postponing reduces
communication overhead only if the samples of one handle are pulled in groups smaller than
$N$ and there is some time between these partial pulls for the parameter
relocation. Assuming that $N$ is bounded from above, it provides \longterm{}. It
does not provide \bounded{} because sampling probabilities depend on the current
allocation of a key (i.e., keys can be postponed to a later sample if they are not local).

\textbf{Local sampling (\nonconform{}). } \sys{} implements local sampling without
active re-partitioning. Instead, \sys{} relies on the application to relocate
parameters: in a relocation PS, the local partition usually
changes constantly, as workers relocate the parameters that they work with (in
direct access). The effect of this local sampling variant heavily depends on
the relocations of the application. Generally, this approach cannot give any
guarantees, as, for example, an application might not relocate parameters at
all. Thus, it generally falls into the \nonconform{} level. In an ideal setting,
however, this approach could provide \longterm{}. For example, this can be the case if an
application partitions its training data randomly and continuously relocates all
parameters (such that a parameter is equally likely to be on all nodes) and
samples uniformly (such that $\pi_k \ll \frac{1}{Q}$ for all $k$).
To make local sampling efficient, \sys{} employs a fast sampling
implementation that does not sample independently.


%% file: tab_sampling_management_techniques.tex
\newcommand{\yes}{\checkmark}
\newcommand{\no}{\boldmath$\times$}

\begin{table}
  \caption{Conformity levels of common sampling schemes.}
  \label{tab:sampling:management-techniques}
  \centering
  \begin{threeparttable}
    \begin{tabular}{lcccc}

      \toprule
          & L1 & L2 & L3 \\
          & \conform{} & \bounded{} & \longterm{} \\
      \midrule
      Independent sampling & \yes & \yes & \yes \\
      Sample reuse & \no & \yes & \yes \\
      Local sampling & \no & \no & \no \\
      Direct-access repurposing & \no & \no & \no \\
      \bottomrule
    \end{tabular}
  \end{threeparttable}
\end{table}


%% file: fig_prep_pull.tex
\newcommand{\atechnique}[5]{
  \addtocounter{numtechniques}{1}

	\node [align=center] at (\thenumtechniques*3.0, 7.1) {#1};
	\node [align=center] at (\thenumtechniques*3.0, 6.4) {\texttt{(#2)}};

	\node (prep\thenumtechniques) [box, draw=primary, fill=primary!5, minimum height=1.8cm] at (\thenumtechniques*3.0, 3.3) {#3};
	\node (pull\thenumtechniques) [box, draw=secondary, fill=secondary!5, minimum height=2.2cm] at (\thenumtechniques*3.0, 0.8) {#4};
	\draw [arrow] (prep\thenumtechniques) to (pull\thenumtechniques); 

	\node (bg\thenumtechniques) [box, draw=bg, fill=bg!5, minimum height=1.5cm] at (\thenumtechniques*3.0, 5.3) {#5};
}

\begin{figure}
  \centering
  \resizebox{1.0\columnwidth}{!}{
    \newcounter{numtechniques}
    \setcounter{numtechniques}{0}
    \begin{tikzpicture}[line width = 0.3mm,
      box/.style={draw, align=center, minimum height=2.2cm, minimum width=2.8cm},
      arrow/.style={->,>=stealth', line width=0.4mm, shorten <=1pt, shorten >=1pt}, 
      every node/.style={font=\Large},
      ]
      
      \node [align=center, text=primary] at (0.8, 3.3) {\texttt{Prepare}\\\texttt{Sample}};
      \node [align=center, text=secondary] at (0.8, 0.8) {\texttt{Pull}\\\texttt{Sample}};
      \node [align=center, text=bg] at (0.8, 5.3) {Back-\\ground\\thread};

      \atechnique{\textbf{Independent}\\\textbf{sampling}}{\conform{}}
      {sample iid. from\\$\pi$ and localize\\(async)}
      {pull parameters \\(remotely\\if necessary)}

      \atechnique{\textbf{Sample}\\\textbf{reuse}}{\bounded{}}
      {re-localize if \\necessary\\(async)}
      {pull parameters \\(remotely\\if necessary)}
      {fill pool:\\sample iid.~from\\$\pi$ and localize}

      \atechnique{\textbf{Sample reuse}\\\textbf{with postponing}}{\longterm{}}
      {re-localize if \\necessary\\(async)}
      {pull parameters\\if local, o/w\\postpone the\\sample}
      {fill pool:\\sample iid.~from\\$\pi$ and localize}

      \atechnique{\textbf{Local}\\\textbf{sampling}}{\nonconform{}}
      {}
      {sample from\\locally available\\part of $\pi$\\and pull locally}
      
    \end{tikzpicture}
  }
  \caption{Sampling scheme implementations in \sys{}.}
  \label{fig:prep_pull}
  \figurespace{}
  \vspace{-0.25cm} 
\end{figure}
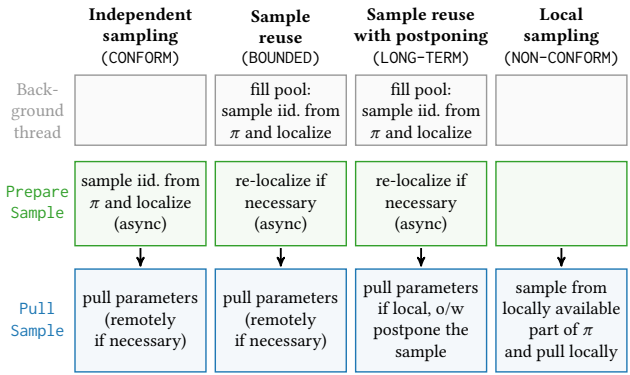


%% file: tab_datasets.tex
\begin{table*}
  \caption{ML tasks, models, datasets, and share of direct and sampling access.}
  \vspace{-0.2cm}
  \label{tab:datasets}
  \centering
  \small
  \begin{threeparttable}
  \begin{tabular}{llrrrlrrrr} \toprule
    Task & \multicolumn{4}{c}{Model parameters} & \multicolumn{3}{c}{Data} & \multicolumn{2}{c}{\hspace{-0.15cm}Parameter access}
    \\
    \cmidrule(lr){2-5} \cmidrule(lr){6-8} \cmidrule(lr){9-10}
         & Model                     & \mbox{Keys}           & \hspace{-1cm}\mbox{Values} & Size & Dataset & \mbox{\hspace{-2.6cm}Data points} & Size & \hspace{-0.3cm}Direct & \hspace{-0.2cm}Sampling 
    \\ \midrule

    Knowledge graph embeddings\hspace{-0.2cm} & \multicolumn{1}{l}{ComplEx, dim.~\num{500}} & \SI{4.8}{M} & \SI{4.8}{B} & \SI{35.9}{GB} & Wikidata5M & \SI{21}{M} & \hspace{-0.3cm}\SI{317}{MB} & 69\% & 31\%
    \\

    Word vectors & \multicolumn{1}{l}{Word2Vec, dim.~\num{1000}} & \SI{1.9}{M} & \SI{1.9}{B} & \SI{7.0}{GB} & 1b word benchmark & \SI{375}{M} & \SI{3}{GB} & 44\% & 56\%
    \\

    Matrix factorization            & \multicolumn{1}{l}{Latent Factors, rank~1000} & \SI{11.0}{M}& \SI{11}{B} & \SI{82.0}{GB} & \rectM{} matrix, zipf 1.1 & \SI{1000}{M} & \SI{31}{GB} & 100\% & 0\%
    \\
    \bottomrule
  \end{tabular}
  \end{threeparttable}
\end{table*}


%% file: 05-experiments.tex
\section{Experiments}
\label{sec:experiments}

We conducted an experimental study to investigate whether and to what extent a
non-uniform PS is beneficial for PS performance. Source code, datasets, and
information for reproducibility are available online.\footnote{\url{https://github.com/alexrenz/NuPS}}

In our study, we compared the
performance of \sys{} to several state-of-the-art PSs on three large-scale ML
tasks (Section~\ref{sec:exp:ete}). Further, we conducted an ablation study
(Section~\ref{sec:exp:ablation}), investigated scalability
(Section~\ref{sec:exp:scalability}), evaluated different sampling schemes
(Section~\ref{sec:exp:sampling}), and explored specific components of \sys{}
(Sections~\ref{sec:exp:techniques} and \ref{sec:exp:so}). Our major insights
are: (i) \sys{} was more than an order of magnitude faster than existing PSs.
(ii) \sys{} achieved best performance when it replicated a small fraction of the
model parameters, and relocated all other parameters. (iii) Both sample reuse
and local sampling significantly reduced communication overhead for sampling
access. We conclude that \emph{a non-uniform PS is key for high performance in
  ML tasks with non-uniform parameter access}.

\subsection{Experimental Setup}
\label{sec:experiments:setup}

We considered three popular ML tasks that require long training: knowledge graph
embeddings, word vectors, and matrix factorization. These tasks are
  representatives for shallow models that exhibit sparse and skewed access. The
tasks differ in multiple ways, including the number of parameters, parameter
access distributions, sampling distribution, and frequency of sampling accesses.
Table~\ref{tab:datasets} provides a summary. In the following, we briefly
discuss each task.

\textbf{Knowledge graph embeddings. } Knowledge graph embedding (KGE) models
learn algebraic representations of the entities and relations in a knowledge
graph. For example, these representations have been applied successfully to
infer missing links in knowledge graphs~\cite{kgcompletion}. This task, based on~\cite{analogy}, trains
ComplEx~\cite{complex} (one of the most popular KGE models) embeddings using SGD
with AdaGrad~\cite{adagrad} and negative sampling~\cite{kgetraining, analogy}.
Negative sampling creates sampling access in this task: to generate negative
samples, both the subject and the object entity of a positive
triple are perturbed $n_{\operatorname{neg}}$ times, by drawing random entities from a uniform
distribution over all entities (we used a common setting of $n_{\operatorname{neg}}=100$~\cite{kgetraining}). We used
the Wikidata5M dataset~\cite{wikidata5m}, a real-world knowledge graph with
\num{4818679} entities and \num{828} relations, and a common embedding size of
\num{500}~\cite{kgetraining}. We partitioned the subject--relation--object
  triples of the dataset to the nodes randomly, as done in~\cite{libkge-dist}. We used
LibKGE~\cite{libkge} (commit 3146885) to evaluate models and report the
\emph{mean reciprocal
  rank (filtered)} (MRRF) as metric for model quality.

\textbf{Word vectors. } Word vectors (WV) are a language modeling technique in
natural language processing: each word of a vocabulary is mapped to a vector of
real numbers~\cite{word2vec,glove,elmo}. These vectors are useful as input for
many natural language processing tasks, for example, syntactic
parsing~\cite{syntactic-parsing} or question
answering~\cite{question-answering}. This task, based on \cite{word2vec}, uses SGD and negative sampling to
train the skip-gram Word2Vec~\cite{word2vec} model (dimension~1000) on the One Billion Word
Benchmark~\cite{1b-word} dataset (with stop words of the Gensim~\cite{gensim}
stop word list removed). The negative sampling creates sampling accesses: in
this task, for each word pair, 3 negative samples are drawn from a distribution that is based
on word frequencies (see Section~\ref{sec:bg:sampling}). We used common
model parameters~\cite{word2vec} for window size (5), minimum count (1), and 
frequent word subsampling (0.01). We measured model accuracy using a common analogical
reasoning task of \num{19544} semantic and syntactic questions~\cite{word2vec}.

\textbf{Matrix factorization. } Low-rank matrix factorization (MF) is a common
tool for analyzing and modeling dyadic data, e.g., in collaborative filtering
for recommender systems~\cite{mf-recsys}. This task, based on \cite{dsgdpp}, uses SGD to factorize a
synthetic, zipf-1.1 distributed \rectM{} dataset with 1b revealed cells, modeled
after the Netflix Prize dataset.\footnote{See \url{https://netflixprize.com/}. We use
a synthetic dataset because the largest openly available dataset that we are
aware of is only \SI{7.6}{GB} large.}
Data points were partitioned to nodes by row and to workers within a node by
  column. Each worker visited its data points by column
(to create locality in column parameter accesses), with random order of columns
and of data points within a column. There is no sampling access
in this task. We report the \emph{root mean squared error} (RMSE) on the test
set as metric for model quality.

\input{fig_end_to_end}

\textbf{Baselines. } We compared performance to a classic PS, to Petuum (a
state-of-the-art replication PS), to Lapse (a state-of-the-art relocation PS),
and to a single node implementation. As classic PS, we used Lapse with
relocation disabled, which, according to~\cite{lapse}, provides performance
similar to PS-Lite. We ran both the SSP and ESSP protocols of
Petuum~\cite{petuum}, with different staleness thresholds. Petuum does
not provide KGE or WV implementations. Thus, we implemented the KGE task
described above in Petuum. We used version 1.1 of Petuum. We did not implement specific sampling schemes in
application code, i.e., applications draw independent samples and access them via
direct access. We used a shared memory
implementation with 8 worker threads as single node baseline.

\textbf{Implementation and cluster.} We implemented \sys{} in C++, using ZeroMQ
and Protocol Buffers for communication, based on PS-Lite~\cite{pslite}. We used
a local cluster of up to 16 Lenovo ThinkSystem SR630 computers, running
Ubuntu Linux 20.04, connected with 100~Gbit Infiniband. Each node was equipped
with two Intel Xeon Silver 4216 16-core CPUs, 512~GB of main memory, and one
2~TB D3-S4610 Intel SSD. We compiled code with g++ 9.3.0, except for
Petuum, which we compiled with g++ 7.5.0, as the compilation with g++ 9.3.0
failed. Unless specified otherwise, we used 8 nodes and 8 worker threads per
node. In Lapse and \sys{}, we additionally used 1 server and 3 ZeroMQ I/O
threads per node. In Petuum, we used 4 communication
channels per node. To prevent exploding gradients, we used gradient norm
clipping as suggested in~\cite{gradient-norm-clipping} for replicated parameters
in the WV and MF tasks (clipping updates that exceed the average norm by more
than 2x). In the KGE task, the use of AdaGrad prevented exploding gradients. For
each task, we tuned hyperparameters on the single node and used the best found
hyperparameter setting in all systems and variants.

\textbf{NuPS.} We ran \sys{} in two configurations: (i) a generally applicable
\emph{untuned} configuration that requires no task-specific tuning and (ii) a
task-specific tuned configuration. The untuned configuration employs a
heuristic to decide the management technique for each parameter: it replicates a
parameter if its access frequency exceeds $100$ times the mean access frequency.
This heuristic is computed from dataset frequency statistics. The untuned
configuration further employs sample reuse without postponing (\bounded{}) with
a use frequency of U=16. To indicate the performance potential of task-specific
insights, we included a tuned configuration by informing our configuration
choices with the results of our detail experiments in Sections~\ref{sec:exp:sampling}
and~\ref{sec:exp:techniques}. The tuned configuration for KGE replicates the 900
most frequently accessed keys (the same as the untuned setting), but uses local
sampling (\nonconform{}). The tuned configuration for WV replicates the
\SI{209}{k} most frequently accessed keys (64x more keys than the untuned
configuration), and employs local sampling (\nonconform{}). For MF, the untuned
configuration seemed to be near-optimal, such that we did not add a separate
tuned configuration. Unless mentioned otherwise, we used the settings of the
untuned configuration and a replica staleness threshold of \SI{40}{ms} in all
experiments. Throughout all experiments, we used a pool size of 250 in the
sample reuse scheme.

\textbf{Measures.} Unless noted otherwise, we ran all variants with a fixed \SI{6}{h} time
budget. We measured model quality over time
and over epochs within this time budget (using the quality metrics described
above). We conducted 3 independent runs of each experiment, each starting from a
distinct randomly initialized model, and report the mean. We depict error bars
for model quality and run time; they present the minimum and
maximum measurements. In some experiments, error bars are not clearly visible
because of small variance. Gray dotted lines indicate the performance of the
single node baseline. Gray shading indicates performance that is dominated by
the single node baseline. We report two types of speedups:
(i) \emph{raw speedup} depicts the speedup in epoch run time, without
considering model quality; (ii) \emph{effective speedup} is calculated from the
time that each variant took to reach 90\% of the best model quality that
the single node baseline achieved. Unless specified otherwise, we report
effective speedups.

\subsection{Overall Performance}
\label{sec:exp:ete}

We investigated the overall effect of a non-uniform PS on PS performance. To do
so, we compared the performance of \sys{} to existing PSs and to the single node
baseline. We ran each variant for the fixed time budget and measured model
quality over this time. Figures~\ref{fig:ete:kge}, \ref{fig:ete:wv}, and
\ref{fig:ete:mf} show model quality over time,
Figures~\ref{fig:ete:kge:epoch}, \ref{fig:ete:wv:epoch}, and
\ref{fig:ete:mf:epoch} show model quality over epoch. \textbf{In summary,
  \sys{} was 31--36x faster than a state-of-the-art replication PS
  (Petuum), 6--46x faster than a state-of-the-art relocation PS (Lapse),
  and 2.3--10.3x faster than the single node baseline.\footnote{The
    comparisons to Petuum and Lapse report raw speedups, because Petuum and
    Lapse did not reach the 90\% thresholds within the time budget.}}

The classic PS was inefficient (with epochs over 7x slower than the single node)
because it accesses parameters over the network, which induced significant
access latency. Lapse was faster than Classic, but still slower than the single
node, because Lapse relocates all parameters, including hot spots. Hot spot
parameters, however, are frequently accessed by multiple nodes simultaneously~\cite{lapse-demo},
such that some of these nodes had to wait for the relocation to finish or access
the parameter remotely, which induced access latency. The per-epoch model
quality of Classic and Lapse was indistinguishable from the single node in KGE
and WV, as these systems provide sequential consistency for all parameters and
employ no specialized sampling schemes. In MF, all distributed variants provided
lower per-epoch quality than the single node, an effect that has been observed
before~\cite{mf-kais}. The step pattern that is visible in MF training stems
from the bold driver heuristic~\cite{bold-driver} that the MF
implementation~\cite{mf-kais} that we adapted uses to tune the learning rate.

For KGE, we ran Petuum SSP and ESSP with staleness thresholds 1, 10, 100, 200, or 1000,
and tried different frequencies for advancing the clock.\footnote{We tried to
  advance the clock after every 1, every 10, and every 100 data points. We
  observed best performance for clocking after every 10th data point. Due to the
  high run times of Petuum, we ran each configuration only once.} None of the
configurations completed the first epoch within the time budget of 6 hours. We
observed the best performance for ESSP with staleness~10, which finished the
first epoch after 13h with a model quality (MRRF) of 0.11. The best SSP run
(staleness 200) finished the first epoch after 15h with a model quality of 0.10.
The reasons for this performance are that Petuum is inefficient
for long tail parameters (as discussed in Section~\ref{sec:skew:sota}) and that
Petuum's replica approach is inefficient for sampling because sampling
access provides no locality: SSP replicas are mostly cold, ESSP
over-communicates. Petuum's MF implementation ran out of memory, because it
stores the training matrix in dense format.
\input{fig_ablation}

The untuned \sys{} configuration outperformed existing PSs across all three
tasks. For KGE and MF, it was also clearly faster than the single node, with up
to 6.7x effective speedups over the single node and minimal negative
effect on (per-epoch) model quality. For WV, however, it barely outperformed the
single node (but still outperformed existing PSs). In contrast, the tuned
configuration provided 4.6--10.3x effective speedups over the single node
across all three tasks. For KGE, the tuned configuration of \sys{}
  provided better per-epoch convergence than the single node. This was an effect
  of local sampling; see Section~\ref{sec:exp:sampling} for more details.

\subsection{Ablation}
\label{sec:exp:ablation}

\sys{} introduces two novel features compared to existing PSs: (i) multi-technique parameter
management and (ii) the integration of sampling into the PS. To investigate
individual effects, we enabled each feature individually and
measured model quality within the time budget. Figure~\ref{fig:abl} shows
the results. We omit MF because there is
no sampling access in MF, such that the entire performance improvement stems
from multi-technique parameter management (which is visible in
Figure~\ref{fig:ete:mf}). \textbf{We found that both multi-technique parameter
management and sampling integration can be beneficial individually, and the
individual benefits compounded when both were combined.}

We compared the performance of four variants: (i) \emph{Lapse}, a relocation PS
without sampling integration; (ii) \emph{Relocation + Replication}, a PS with
multi-technique parameter management but without sampling integration; (iii)
\emph{Relocation + Sampling}, a relocation-only PS with sampling integration;
(iv) \emph{\sys{}}, a multi-technique PS with sampling integration. Going from a single-technique
relocation PS to a multi-technique PS made an epoch 67--73\% faster with only 
small effect on model quality. Adding sampling support to the relocation PS
made an epoch 17--62\% faster, with a small negative effect on model
quality. The combination of both made an epoch 94\% faster, with a small
negative effect on per-epoch model quality.

\subsection{Scalability}
\label{sec:exp:scalability}
\input{fig_scale}
\input{fig_scale_effective}

To investigate scalability, we ran Lapse, the best Petuum SSP and ESSP
  configurations, and \sys{} for one epoch on 1, 2, 4, 8, and 16 nodes and
  calculated the raw speedup. Figure~\ref{fig:scale} depicts the results.
  Further, we ran convergence experiments on 16 nodes for those systems that
  reached the 90\% model quality threshold on 8 nodes.
  Figure~\ref{fig:eff-scale} depicts the effective speedup for these systems.
  \textbf{Overall, \sys{} scaled more efficiently than other PSs, with up to
    near-linear raw and up to superlinear effective speedups.}

We first discuss raw scalability, i.e., the speedup w.r.t. epoch run
  time (Figure~\ref{fig:scale}). On a single
  node, \sys{} and Lapse were faster than Petuum because \sys{} and Lapse access
  local parameters via shared memory, whereas Petuum sends intra-process
  messages to do so. Lapse provided poor scalability because the more nodes are
  used, the higher the chance that multiple nodes access a parameter at the same
  time and, thus, that they have to wait for a relocation to finish or to access
  parameters remotely.  
  Neither Petuum ESSP nor SSP outperformed the shared-memory single-node
  baseline, even on 16~nodes. ESSP scaled poorly even when compared to its own
  (inefficient) run time on a single node (4.8x faster on 16 nodes)
  because its eager replication protocol over-communicates: after a short
  warm-up period, each node holds a replica of the full model. The more nodes,
  the more replicas had to be synchronized, such that replica synchronization
  became a bottleneck. The lazy replication protocol of SSP scaled better than ESSP
  compared to its own (inefficient) run time on a single node (12x faster
  on 16 nodes), but its overall performance was poor because its replicas
  were cold most of the time (and thus required synchronous replica
  refreshes).

\sys{} scaled more efficiently than existing PSs because it (i) limits the
  bottleneck of eager replication by replicating only a small subset of hot spot
  parameters, (ii) prevents the majority of relocation conflicts by employing
  relocation only for long tail parameters, and (iii) employs sampling schemes
  to reduce sampling communication overhead. With 16 nodes, it provided up
  to 13.4x raw speedups over the shared memory single node. \sys{}
  further provided up to 20x effective speedups for KGE and 8x for MF (see
  Figure~\ref{fig:eff-scale}). For WV, although the raw speedup on 16 nodes was
  10.2x, the effective speedup was only 2.2x. The reason for this is
  that we used the hyperparameter configuration that worked best on the single
  node throughout all experiments. With other
  hyperparameters, we observed better effective speedups for WV.

\input{fig_scl}

\subsection{Effect of Sampling Schemes}
\label{sec:exp:sampling}

We investigated the effect of different sampling schemes in
\sys{} on run time and model quality. To do so, we ran KGE and WV with different
sampling schemes: independent sampling (\conform{}), U=16 and U=64 sample
reuse without postponing (\bounded{}) and with postponing
(\longterm{}), and local sampling (\nonconform{}). Figures~\ref{fig:scl:kge} and
\ref{fig:scl:wv} show model quality over time, 
Figures~\ref{fig:scl:kge:epoch} and \ref{fig:scl:wv:epoch} show model quality
over epoch. We omit MF as it does not contain sampling access. We
further omit the results from sample reuse with postponing as its results were
within 10\% of sample reuse without postponing.\footnote{Postponing made no
  measurable difference in KGE, and sped up WV run times by 10\%, with no
  measurable impact on model quality.} \textbf{We
found that both sample reuse and local sampling led to significant speedups
over independent sampling, with small negative or---in the case of local sampling---even
positive effects on per-epoch model quality.}

\emph{Independent sampling} provided per-epoch quality near-identical to the single node,
but was slowest, because it induced high communication overhead for each sample.
\emph{Sample reuse} had lower communication overhead (and, thus, faster epoch
run times), but at the cost of a (small) negative effect on per-epoch model quality. The
higher the use frequency, the faster an epoch and
the larger the negative effect on quality. The U=16 variant provided a good
compromise, with minimal effect on model quality and fast run times.

\emph{Local sampling} exhibited excellent performance,
despite providing no guarantees on sampling quality: it was fast and per-epoch
model quality was as good as the single node in WV, and was
\emph{better} in KGE. We hypothesize that this mainly is because \sys{} combines local sampling with dynamic
allocation: both tasks continuously relocate model parameters, such
that the local parameter partitions contain many different parameters over
time. To evaluate this hypothesis, we ran local sampling with \emph{static
  allocation} in KGE.
Figure~\ref{fig:scl:kge:epoch} includes the results: with static
allocation, model quality deteriorated drastically. We further conjecture
  that the reason for the better-than-single-node quality of local sampling in
  KGE was that relocation led to local samples that were more
  informative than global samples. Similar effects have been observed
  previously~\cite{dglke}.

\subsection{Choice of Management Technique}
\label{sec:exp:techniques}
\input{fig_nhs}

\input{tab_replica_stats}
\input{fig_sync_overhead}

We investigated how the choice of management technique, i.e., the choice of
whether to replicate or relocate a key, affects the performance of \sys{}. The
\sys{} untuned heuristic replicates the \num{900} most frequent keys in KGE, the
\num{3272} most frequent keys in WV, and the \num{755} most frequent column
keys in MF. We varied these numbers by factors $\frac{1}{64}$, $\frac{1}{16}$,
$\frac{1}{4}$, $4$, $16$, $64$, and $256$. The leftmost columns of
Table~\ref{tab:replica-stats} depict what share of keys was replicated for each
setting. We ran one epoch of each setting and measured epoch run time
and model quality. Figure~\ref{fig:nhs} depicts the results. \textbf{We found that
  it was crucial for performance to replicate ``enough'' parameters such that
  the set of hot spot parameters is managed by replication, but not too many
  parameters, as replication created significant over-communication for long
  tail parameters.}

This effect was visible for all tasks: starting from no replicated keys (i.e.,
all keys managed by relocation), increasing the number of replicated keys first
improved run time, and had minimal effect on model quality. However, after some
point, replicating more keys deteriorated model quality, and even slowed down
run time for KGE and MF. The reason for the negative effect on model quality was
that the replicas were stale, because the replica updates became too large to
synchronize them frequently over the network of the cluster. We
configured \sys{} to provide the default \SI{40}{ms} staleness bound (i.e., 25
synchronizations per second), but to \emph{not} block operations when it did not
reach this goal. Figure~\ref{fig:nhs} includes the actual synchronization
frequency if model quality was not within 10\% of the model quality without
replication. The middle columns of Table~\ref{tab:replica-stats} provide the
size of the replicated values for all settings. For example, the 64x WV setting
replicated \SI{799}{MB} of parameter values. Large numbers of replicated keys led to slower
epoch run times for KGE and MF, because relocation operations competed with
replica synchronization for network bandwidth. This effect was not visible for
WV because, in WV, the majority of accesses went to replicated keys (and, thus,
were fast despite network congestion). The share of accesses that went to
replicas is depicted in the rightmost columns of Table~\ref{tab:replica-stats}.
For example, 88\% of all accesses went to replicas in the 64x WV setting. 

\subsection{Effect of Replica Staleness}
\label{sec:exp:so}

We investigated the effect of replica staleness on epoch run time and model
quality. To do so, we varied synchronization frequency: we synchronized replicas
either 125, 25, 5, 1, or 0.2 times per second or not at all. We ran one epoch of each
setting and measured epoch run time and model quality after this epoch. Note
that without replica synchronization, nodes may hold different models. In these
cases, we evaluated the model of the first node. Figure~\ref{fig:sync-overhead}
reports the results. \textbf{Overall, replication had only minimal effect on
  model quality when replica staleness was low.}

Replication had only small effect on model quality when replicas were
synchronized at least 5 times per second. In contrast, infrequent synchronization
(less than once per second) deteriorated model quality drastically in KGE and
WV. However, infrequent synchronization (or no synchronization at all) worked
well in some settings (in particular in MF). We theorize that the reason for
this was that \sys{} employs replication for only a small subset of parameters,
such that replication parameters are kept synchronized indirectly through the
parameters that are managed by relocation.

\subsection{Comparison to Task-Specific Implementations}
\label{sec:exp:task-specific}
In a general-purpose system, a performance overhead over optimized
  task-specific implementations is expected. 
  To investigate the extent of this overhead in \sys{}, we compared to
  specific implementations for each task. Each of these implementations is
  specialized and highly tuned for the respective task. In contrast to a
  general-purpose PS such as \sys{}, these implementations cannot be used to run
  other ML tasks. Note that some of these implementations use different, more
  complex training algorithms than the implementations in \sys{}.
  \textbf{Overall, we found that \sys{} was competitive to specialized and tuned
    task-specific implementations.}

For MF, we compared to the highly tuned MPI implementations of DSGD and
  DSGD++~\cite{dsgdpp}. We ran convergence experiments on 8
  and 16 nodes. We measured how long the implementations took to reach the 90\% quality
  threshold. We used the same
  hyperparameters, model starting points, and learning rate schedule across
  DSGD, DSGD++, and \sys{}. On 8 nodes, \sys{} was 16\% faster than DSGD and 15\%
  slower than DSGD++. On 16 nodes, \sys{} was 37\% faster than DSGD and 16\%
  faster than DSGD++.

  For KGE, we compared to the highly specialized framework
  PyTorch-BigGraph~\cite{biggraph}. Note that PyTorch-BigGraph is designed for a
  different training algorithm, with different hyperparameters: to reduce
  communication overhead, it uses mini-batch SGD, whereas the KGE implementation
  in \sys{} employs regular SGD (i.e., batch size 1). To minimize the impact of
  algorithm hyperparameters in our comparison, we compared epoch run times.
  \sys{} ran an epoch in 12 minutes on 16 nodes (24 minutes on 8 nodes). In this
  setting (i.e., batch size 1), PyTorch-BigGraph was much slower than \sys{}: it
  took more than 5 hours to run one epoch, both on 8 and 16 nodes. Using a very
  large batch size led to faster epochs in PyTorch-BigGraph (up to 3x faster
  with batch size 1000 than \sys{} with batch size 1), but can also be
  implemented in \sys{}.

For WV, we are not aware of a highly tuned and publicly available
  distributed implementation, so we compared to two highly tuned single-node
  implementations: the original C implementation of Word2Vec~\cite{word2vec} and
  Gensim~\cite{gensim}. The implementation in Gensim and the one in \sys{} are
  both based on the original C implementation. For both single-node
  implementations, we achieved the fastest epoch run times with 64 threads.
  Gensim completed an epoch in 15 minutes, the original implementation in 12
  minutes. With 8x8 threads, \sys{} took 13.5 minutes for one epoch; with 16x8
  threads, it took 8 minutes. One factor that limits the performance of \sys{}
  compared to these task-specific implementations is that---as other general-purpose
  PSs~\cite{lapse}---\sys{} provides per-key atomic updates. To achieve this,
  workers receive dedicated working copies of parameters. Creating these copies
  and writing updates back into the parameter store creates overhead compared to
  the task-specific WV implementations, which let workers read and write in the
  parameter store directly, without any consistency or isolation guarantees.
  Empirically, this works well for this particular task, but the effects for
  other tasks in a general-purpose system are unclear.


%% file: fig_end_to_end.tex
\begin{figure*}
  \centering
  \includegraphics[page=1,width=1.0\textwidth]{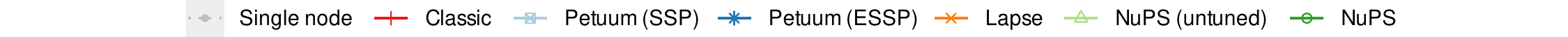}

  \begin{subfigure}[b]{0.33\textwidth}
    \centering
    \includegraphics[page=1,width=0.95\textwidth]{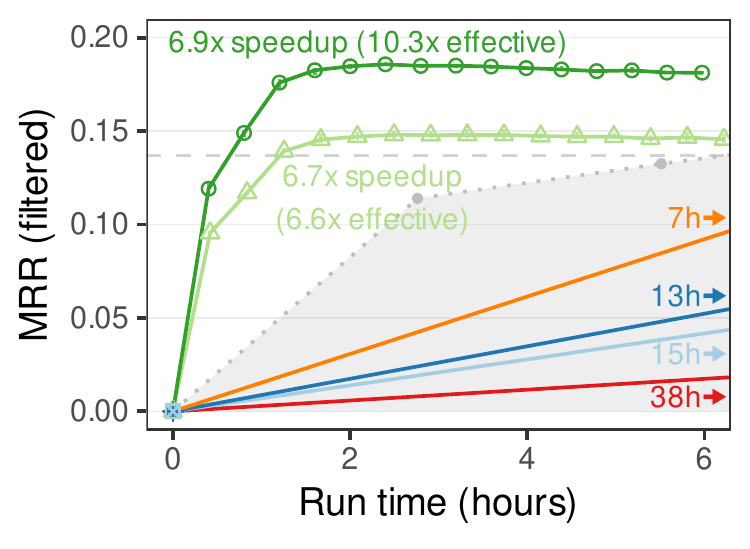}
    \caption{KGE (quality over time)}
    \label{fig:ete:kge}
  \end{subfigure}%
  \begin{subfigure}[b]{0.33\textwidth}
    \centering
    \includegraphics[page=1,width=0.95\textwidth]{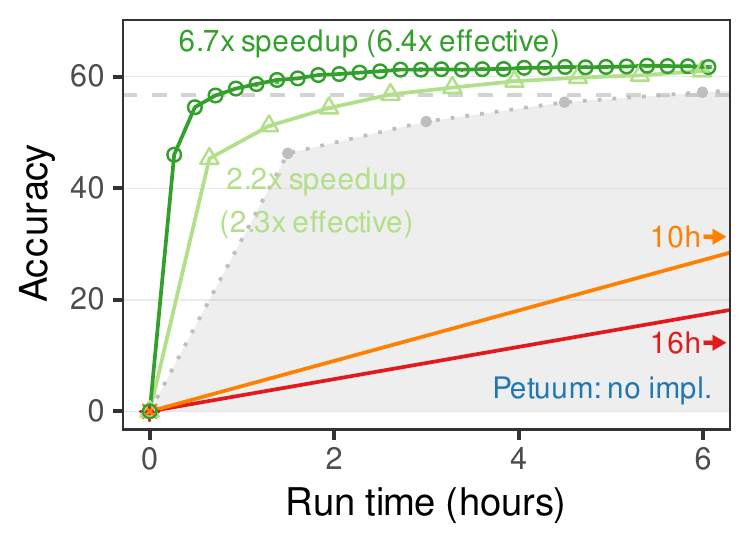}
    \caption{WV (quality over time)}
    \label{fig:ete:wv}
  \end{subfigure}%
  \begin{subfigure}[b]{0.33\textwidth}
    \centering
    \includegraphics[page=1,width=0.95\textwidth]{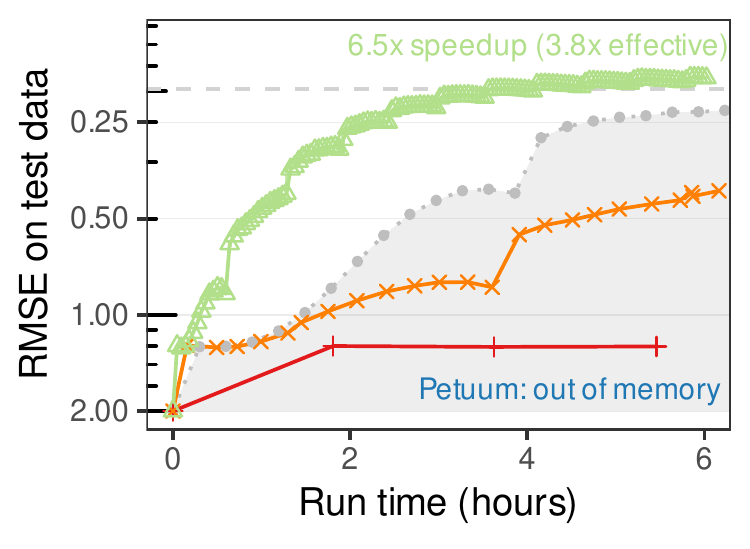}
    \caption{MF (quality over time)}
    \label{fig:ete:mf}
  \end{subfigure}%
  \vspace{0.1cm}

  \begin{subfigure}[b]{0.33\textwidth}
    \centering
    \includegraphics[page=1,width=0.95\textwidth]{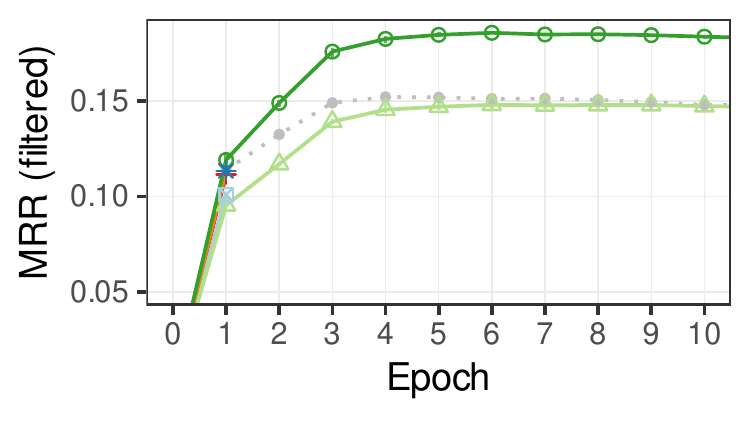}
    \caption{KGE (quality over epoch)}
    \label{fig:ete:kge:epoch}
  \end{subfigure}%
  \begin{subfigure}[b]{0.33\textwidth}
    \centering
    \includegraphics[page=1,width=0.95\textwidth]{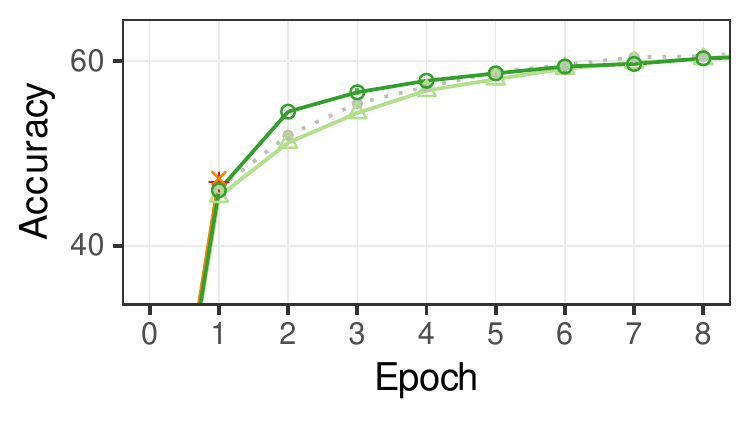}
    \caption{WV (quality over epoch)}
    \label{fig:ete:wv:epoch}
  \end{subfigure}%
  \begin{subfigure}[b]{0.33\textwidth}
    \centering
    \includegraphics[page=1,width=0.95\textwidth]{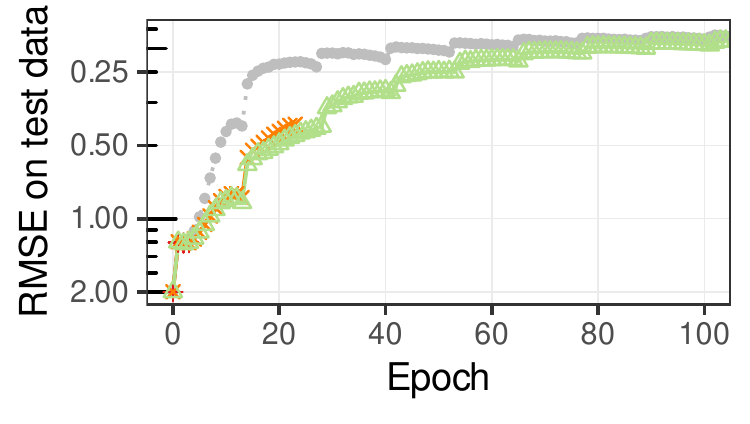}
    \caption{MF (quality over epoch)}
    \label{fig:ete:mf:epoch}
  \end{subfigure}%
  \caption{End-to-end performance of different PSs on 8 nodes. \sys{}
      outperformed Petuum (a state-of-the-art replication PS) and Lapse (a
      state-of-the-art replication PS) by up to one order of magnitude and
      provided up to linear scalability over the single node. The gray shaded
      area indicates performance that is dominated by the single node. Error
      bars depict minimum and maximum measurements for run time and model
      quality (but are often not visible due to low variance). The dashed gray
      line depicts the model quality threshold at which effective speedups are
      computed (90\% of the best observed single-node quality).}
  \label{fig:ete}
  \figurespace{}
\end{figure*}


%% file: fig_ablation.tex
\begin{figure}
  \centering
  \includegraphics[page=1,width=1.0\columnwidth]{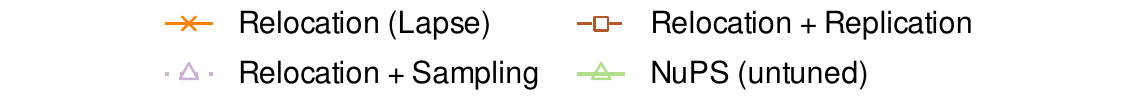}

  \begin{subfigure}[b]{.5\columnwidth}
    \centering
    \includegraphics[page=1,width=1.0\columnwidth]{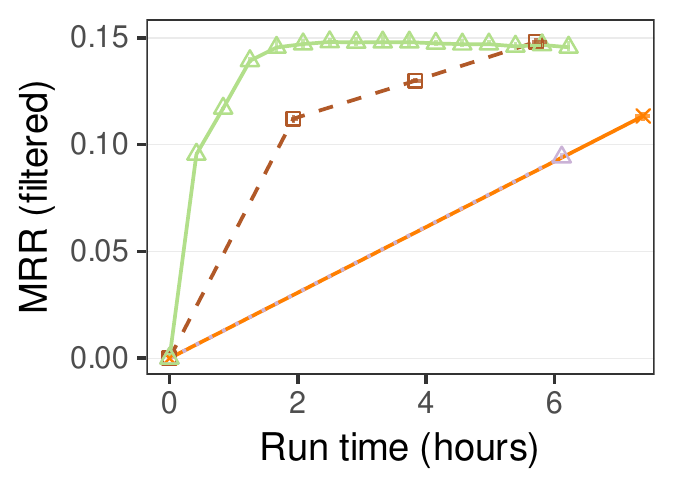}
    \caption{Ablation, KGE}
    \label{fig:abl:kge}
  \end{subfigure}%
  \begin{subfigure}[b]{.5\columnwidth}
    \centering
    \includegraphics[page=1,width=1.0\columnwidth]{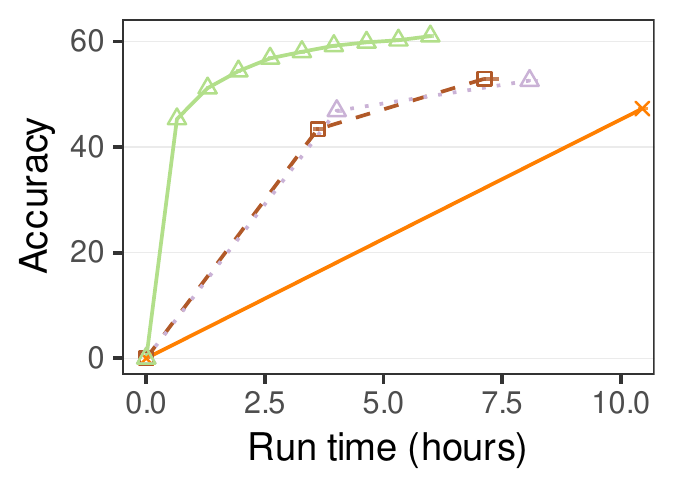}
    \caption{Ablation, WV}
    \label{fig:abl:wv}
  \end{subfigure}%
  \captionsetup{skip=6pt} 
  \caption{Ablation. Both (i) combining replication and relocation and
    (ii) integrating specialized sampling access management techniques improved performance
    individually, and it was beneficial to combine the two. }
  \label{fig:abl}
  \figurespace{}
  \vspace{-0.2cm} 
\end{figure}


%% file: fig_scale.tex
\begin{figure}
  \includegraphics[page=1,width=1.0\columnwidth]{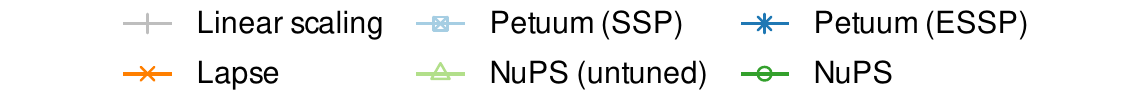}

  \begin{subfigure}[b]{.35\columnwidth}
    \centering
    \includegraphics[page=1,width=1.0\textwidth]{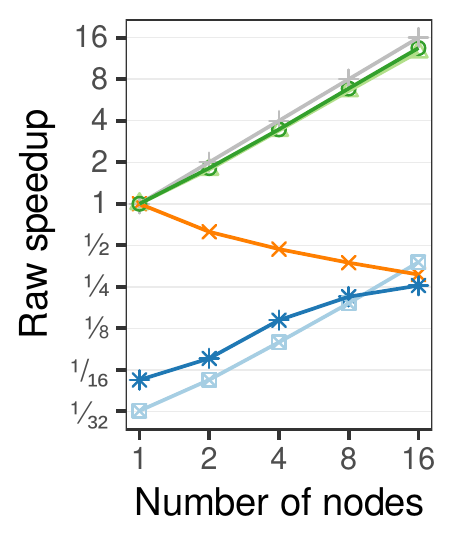}
    \caption{KGE}
    \label{fig:scale:kge}
  \end{subfigure}%
  \begin{subfigure}[b]{.324\columnwidth}
    \centering
    \includegraphics[page=1,width=1.0\textwidth]{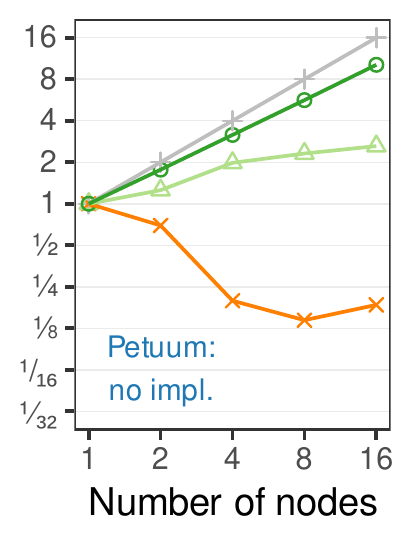}
    \caption{WV}
    \label{fig:scale:wv}
  \end{subfigure}%
  \begin{subfigure}[b]{.324\columnwidth}
    \centering
    \includegraphics[page=1,width=1.0\textwidth]{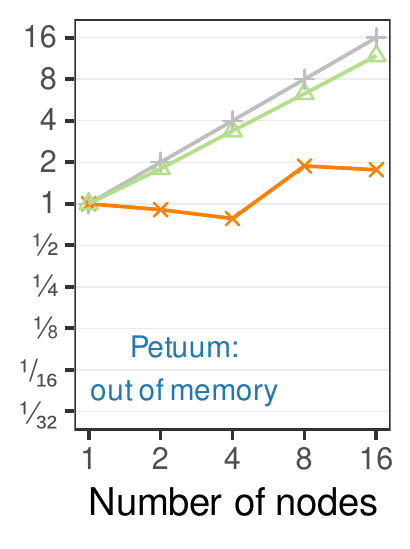}
    \caption{MF}
    \label{fig:scale:mf}
  \end{subfigure}%
  
  \caption{Raw scalability (logarithmic axes). The y-axis depicts raw
    speedup, i.e., speedup w.r.t. epoch run time over the shared-memory
    single-node baseline. \sys{} scaled more efficiently than other PSs,
    with up to near-linear speedups.}
  \label{fig:scale}
  \figurespace{}
\end{figure}


%% file: fig_scale_effective.tex
\begin{figure}
  \includegraphics[page=1,width=1.0\columnwidth]{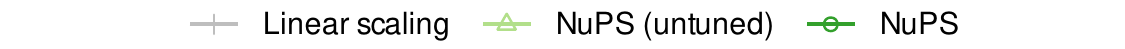}

  \begin{subfigure}[b]{.35\columnwidth}
    \centering
    \includegraphics[page=1,width=1.0\textwidth]{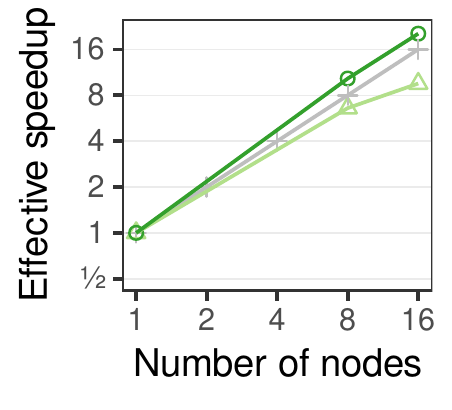}
    \caption{KGE}
    \label{fig:eff-scale:kge}
  \end{subfigure}%
  \begin{subfigure}[b]{.324\columnwidth}
    \centering
    \includegraphics[page=1,width=1.0\textwidth]{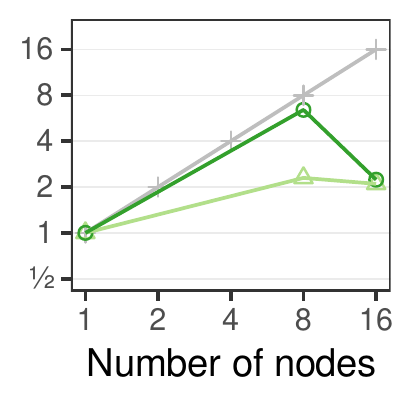}
    \caption{WV}
    \label{fig:eff-scale:wv}
  \end{subfigure}%
  \begin{subfigure}[b]{.324\columnwidth}
    \centering
    \includegraphics[page=1,width=1.0\textwidth]{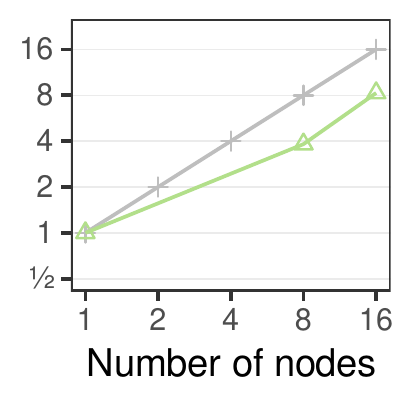}
    \caption{MF}
    \label{fig:eff-scale:mf}
  \end{subfigure}%
  
  \caption{Effective scalability (logarithmic axes). The y-axis depicts
    effective speedup, i.e., speedup w.r.t. reaching 90\% of the best
  model quality observed on a single node.}
  \label{fig:eff-scale}
  \figurespace{}
\end{figure}


%% file: fig_scl.tex
\begin{figure}
  \includegraphics[page=1,width=1.0\columnwidth]{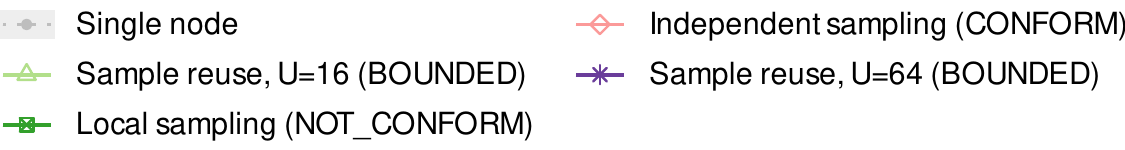}

  \begin{subfigure}[b]{.5\columnwidth}
    \centering
    \includegraphics[page=1,width=1.0\textwidth]{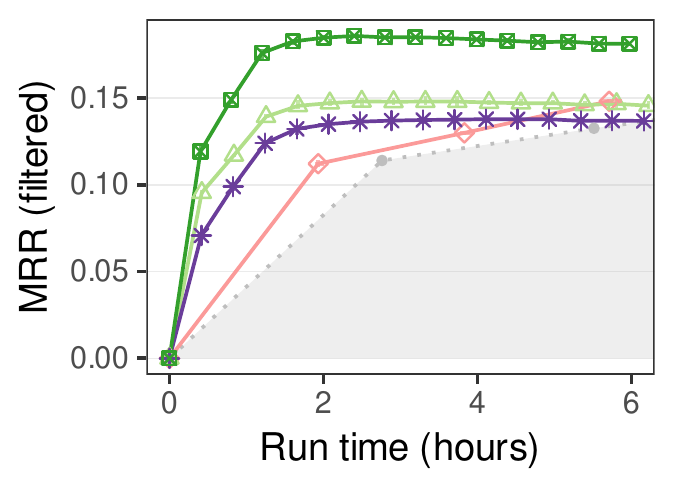}
    \caption{KGE (quality over time)}
    \label{fig:scl:kge}
  \end{subfigure}%
  \begin{subfigure}[b]{.5\columnwidth}
    \centering
    \includegraphics[page=1,width=1.0\textwidth]{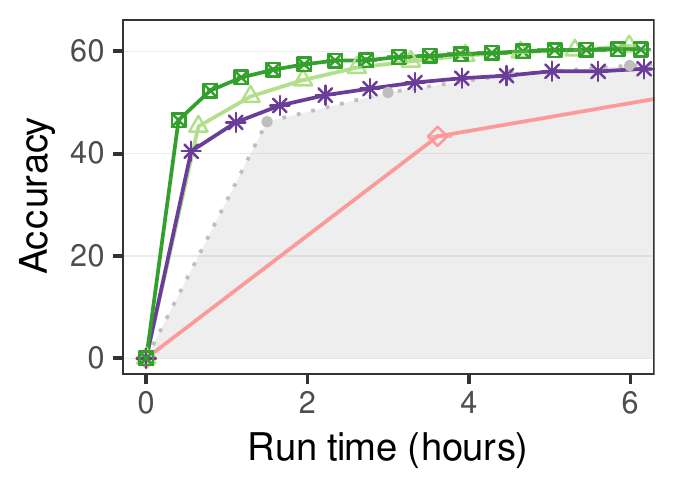}
    \caption{WV (quality over time)}
    \label{fig:scl:wv}
  \end{subfigure}

  \vspace{0.1cm}

  \begin{subfigure}[b]{.5\columnwidth}
    \centering
    \includegraphics[page=1,width=1.0\textwidth]{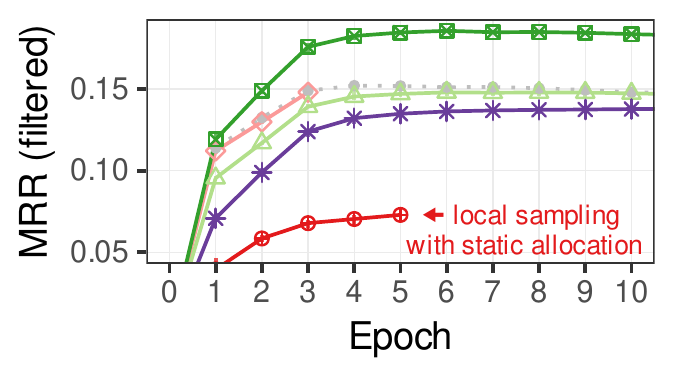}
    \caption{KGE (quality over epoch)}
    \label{fig:scl:kge:epoch}
  \end{subfigure}%
  \begin{subfigure}[b]{.5\columnwidth}
    \centering
    \includegraphics[page=1,width=1.0\textwidth]{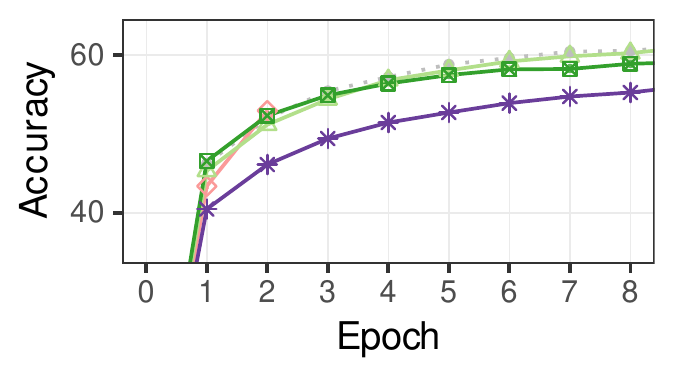}
    \caption{WV (quality over epoch)}
    \label{fig:scl:wv:epoch}
  \end{subfigure}%
  
  \caption{Performance of different sampling access management techniques. Both
    sample reuse and local sampling led to significant speedups over relocation.}
  \label{fig:scl}
  \figurespace{}
\end{figure}


%% file: fig_nhs.tex
\begin{figure*}
  \centering

  \begin{subfigure}[b]{0.38\textwidth}
    \centering
    \includegraphics[page=1,width=1.0\textwidth]{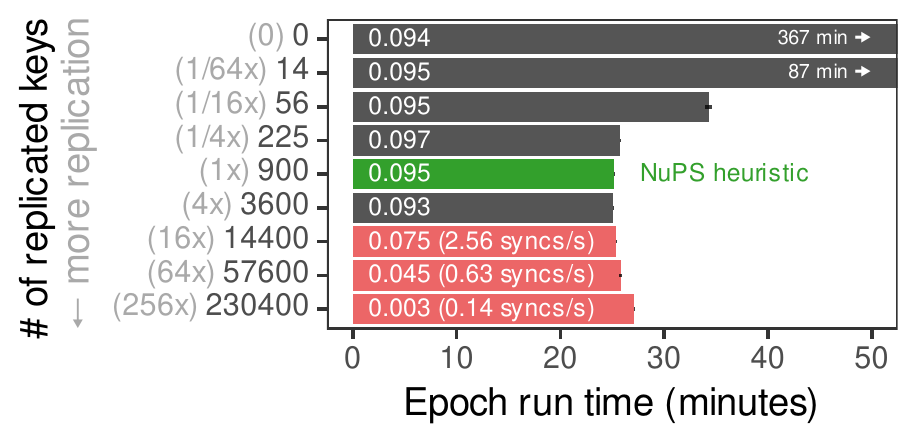}
    \caption{KGE}
    \label{fig:nhs:kge}
  \end{subfigure}%
  \begin{subfigure}[b]{0.31\textwidth}
    \centering
    \includegraphics[page=1,width=1.0\textwidth]{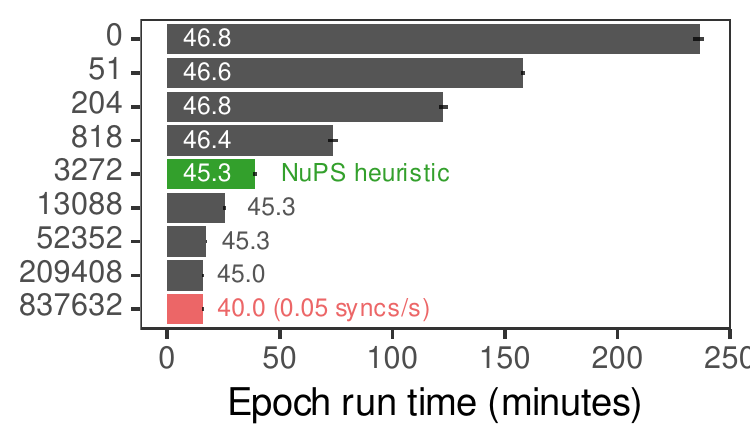}
    \caption{WV}
    \label{fig:nhs:wv}
  \end{subfigure}%
  \begin{subfigure}[b]{0.31\textwidth}
    \centering
    \includegraphics[page=1,width=1.0\textwidth]{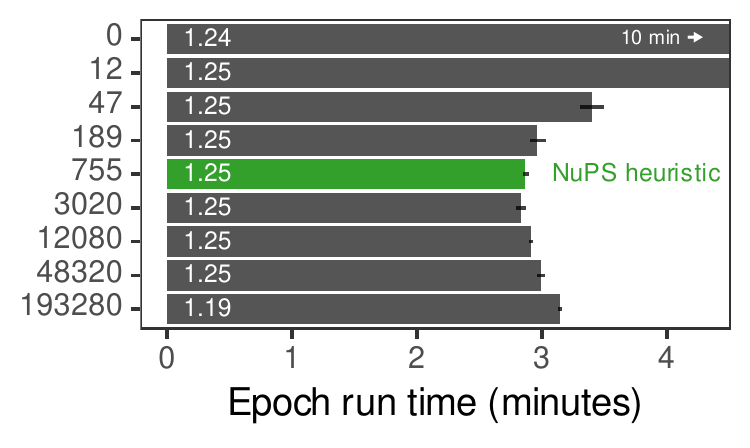}
    \caption{MF}
    \label{fig:nhs:mf}
  \end{subfigure}%
  \caption{Impact of the management technique on epoch run time and model
    quality. The numbers in the plots depict model quality. A run is marked red
    if the resulting model quality was not within 10\% of the model quality
    without replication. For these runs, the numbers in the plots additionally
    depict the actual synchronization frequency.}
  \label{fig:nhs}
  \figurespace{}
\end{figure*}


%% file: tab_replica_stats.tex
\newcommand\eat{\hspace{-0.4cm}}
\begin{table}
  \caption{Share of replicated keys, replica size, and share of accesses to
    replicas for different extents of replication. A cell is marked red if the
    resulting model quality was not within 10\% of the quality without
    replication.}
  \label{tab:replica-stats}
  \centering
  \small
  \begin{threeparttable}
    \addtolength{\tabcolsep}{-0.5pt}
    \begin{tabular}{rrrrrrrrrr} \toprule
      & \multicolumn{3}{c}{\makecell{Replicated \\ keys (\%)}} & \multicolumn{3}{c}{\makecell{Size of replicated \\ values (MB)}} &  \multicolumn{3}{c}{\makecell{Accesses to\\ replicas (\%)}} \\ 
      \cmidrule(lr){2-4} \cmidrule(lr){5-7} \cmidrule(lr){8-10}
         Factor    & KGE & WV & MF & KGE & WV & MF & KGE & \hspace{-0.2cm}WV & \hspace{-0.1cm}MF \\
      \midrule 

0 & \num{0.0000} & \num{0.0000} & \num{0.0000} & \num{0} & \num{0} & \num{0} & \num{0} & \num{0} & \num{0}\\ 
\SI[parse-numbers=false]{1/64}{x} & \num{0.0003} & \num{0.0027} & \num{0.0001} & \num{0} & \num{0} & \num{0} & \num{23} & \num{7} & \num{3}\\ 
\SI[parse-numbers=false]{1/16}{x} & \num{0.0012} & \num{0.0108} & \num{0.0004} & \num{0} & \num{1} & \num{0} & \num{33} & \num{13} & \num{5}\\ 
\SI[parse-numbers=false]{1/4}{x} & \num{0.0047} & \num{0.0435} & \num{0.0017} & \num{2} & \num{3} & \num{1} & \num{38} & \num{25} & \num{9}\\ 

\midrule \SI{1}{x} & \multirow{2}{*}{\num{0.0187}} & \multirow{2}{*}{\num{0.1740}} & \multirow{2}{*}{\num{0.0069}} & \multirow{2}{*}{\num{7}} & \multirow{2}{*}{\num{12}} & \multirow{2}{*}{\num{6}} & \multirow{2}{*}{\num{41}} & \multirow{2}{*}{\num{45}} & \multirow{2}{*}{\num{14}}\\
\hspace{-0.15cm} {\footnotesize (heuristic)}\hspace{-0.15cm} \\

\midrule \SI{4}{x} & \num{0.0747} & \num{0.6958} & \num{0.0275} & \num{27} & \num{50} & \num{23} & \num{44} & \num{67} & \num{19}\\ 
\SI{16}{x} & \bad{\num{0.2988}} & \num{2.7832} & \num{0.1098} & \bad{\num{110}} & \num{200} & \num{92} & \bad{\num{45}} & \num{82} & \num{24}\\ 
\SI{64}{x} & \bad{\num{1.1951}} & \eat\num{11.1330} & \num{0.4393} & \bad{\num{439}} & \num{799} & \num{369} & \bad{\num{47}} & \num{88} & \num{30}\\ 
\SI{256}{x} & \bad{\num{4.7806}} & \eat\bad{\num{44.5319}} & \num{1.7571} & \eat\bad{\num{1758}} & \eat\bad{\num{3195}} & \eat\num{1475} & \bad{\num{52}} & \bad{\num{92}} & \num{37}\\ 
      \bottomrule
    \end{tabular}
  \end{threeparttable}
\end{table}


%% file: fig_sync_overhead.tex
\begin{figure*}
  \centering
  \begin{subfigure}[b]{.365\textwidth}
    \centering
    \includegraphics[page=1,width=1.0\textwidth]{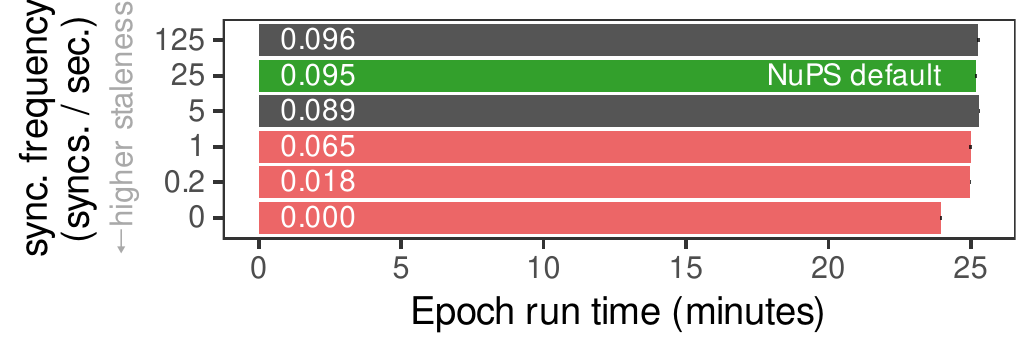}
    \caption{KGE}
    \label{fig:sync-overhead:kge}
  \end{subfigure}%
  \begin{subfigure}[b]{.317\textwidth}
    \centering
    \includegraphics[page=1,width=1.0\textwidth]{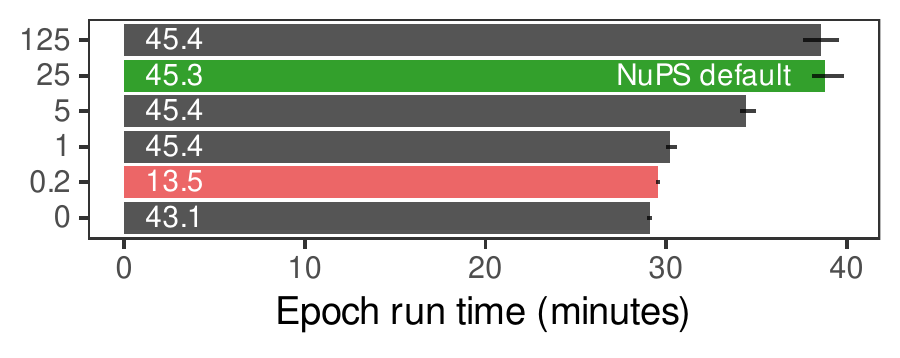}
    \caption{WV}
    \label{fig:sync-overhead:wv}
  \end{subfigure}%
  \begin{subfigure}[b]{.317\textwidth}
    \centering
    \includegraphics[page=1,width=1.0\textwidth]{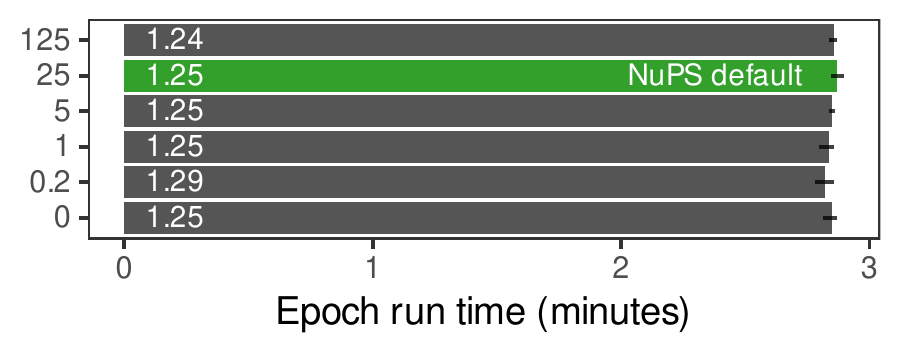}
    \caption{MF}
    \label{fig:sync-overhead:mf}
  \end{subfigure}%
  \caption{Effect of replica staleness on epoch run time and model quality. The
    numbers in the bars depict model quality after one epoch. Bars are marked red
  if model quality was not within 90\% of the quality produced by a
  setting with no replication.}
  \label{fig:sync-overhead}
  \figurespace{}
\end{figure*}


%% file: 06-conclusion.tex
\section{Conclusion}

We explored how to extend the scope of PSs to ML with non-uniform parameter
access. To this end, we presented \sys{}, a non-uniform PS that employs
multi-technique parameter management to efficiently handle skew and integrates
sampling schemes to efficiently handle sampling. We found that a non-uniform
PS can be highly beneficial: in our experimental study,
\sys{} outperformed existing PSs by up to one order of magnitude. These
results open up several research directions for further improving PS
performance: integrating more management techniques (e.g., highly
specialized ones), developing further sampling schemes, and devising
fine-grained methods for picking management techniques for parameters.
